\documentclass[%
reprint,
 superscriptaddress,
 amsmath,amssymb,
 aps,
pra,
floatfix,
longbibliography]{revtex4-1}

\usepackage{graphicx}
\usepackage{subfigure}
\usepackage{dcolumn}
\usepackage{bm}
\usepackage{epstopdf}
\usepackage{xcolor}
\usepackage{hyperref}
\usepackage{appendix} 
\usepackage{mathtools}
\usepackage{tablefootnote}
\usepackage{comment}
\usepackage{textcase}
\usepackage{soul}
\usepackage{float}
\usepackage[normalem]{ulem}
\usepackage[english]{babel}
\usepackage{braket}
\usepackage{multirow}

\def\phm{\phantom{-}}
\def\pp {\phantom{00}}

\begin{document}
\setstcolor{red}

\title{Competing Ionization and Dissociation:  Extension of the energy-dependent frame transformation to the gerade symmetry of H$_2$}

\author{D\'{a}vid Hvizdo\v{s}}
\affiliation{Department of Physics and Astronomy, Purdue University, West Lafayette, Indiana 47907 USA}
\author{Roman \v{C}ur\'{\i}k}
\affiliation{J. Heyrovsk\'{y} Institute of Physical Chemistry, ASCR,
Dolej\v{s}kova 3, 18223 Prague, Czech Republic}
\author{Chris H. Greene}
\affiliation{Department of Physics and Astronomy, Purdue University, West Lafayette, Indiana 47907 USA}
\affiliation{Purdue Quantum Science and Engineering Institute, Purdue University, West Lafayette, Indiana 47907 USA}

\date{\today}

\begin{abstract}
    This article solves two major tasks that frequently arise in the theory of electron collisions with a target molecular cation.  First, it extends the energy-dependent frame transformation treatment(EDFT), which is needed to map fixed-nuclei electron-molecule scattering matrices into an energy-dependent laboratory frame scattering matrix with vibrational channel indices.  The EDFT mapping can now be carried out even when the target molecule possesses multiple low energy potential curves, significantly transcending previous applications.  Secondly, it implements a method to extract the rest of the full lab-frame scattering matrix, i.e. the columns and rows describing input and/or output dissociation channels.
    The treatment is  
    benchmarked in this article against the essentially exact solution of a refined two-dimensional model of the singlet {\it gerade} $\Sigma$ symmetry of H$_2$. Our tests demonstrate that the theory  accurately maps fixed-nuclei scattering information, of the type provided by existing electron-molecule computer codes, into a laboratory-frame scattering matrix that includes both ionization and dissociation.  This treatment can provide a general framework applicable to a broad class of electron collision processes involving diatomic target ions, suitable for an accurate description of challenging processes such as dissociative recombination.
\end{abstract}

\pacs{Valid PACS appear here}

\maketitle

\section{Introduction}
The multichannel quantum defect description of the Rydberg states of molecular hydrogen and its isotopologues has achieved remarkable success in describing the high resolution spectroscopy of H$_2$.  When an extreme ultraviolet (XUV) photon strikes the molecular ground state, it primarily excites the {\it ungerade} states dominated by single electron photoabsorption and, at energies above the lowest ionization threshold, leading primarily to photoionization although at some energies photodissociation can dominate.  The {\it gerade} states are instead most easily probed spectroscopically through two-photon absorption.  

Other physical collision processes important in molecular hydrogen include electron-ion collisions that can change the H$_2^+$ rovibrational level, or where the electron is captured and dissociates the molecule into neutral atoms H$+$H or into ions(p$+$H$^-$). At higher energies 3-body or even 4-body breakup also become possible collisional outcomes.  And of course the time-reverse of all those processes also occur in many natural physical contexts.  This four-particle system with two nuclei and two electrons has extremely rich collision physics and spectroscopy, and it continues to pose serious difficulties for theoretical methods. One of the most interesting energy ranges that has received extensive attention over the past 50 years is the range below the lowest dissociation threshold of the cation, which is the energy of H(1s) + p + e with no kinetic energy in the system.  In that energy range, only two-body fragmentation channels are energetically open and can serve as collision entrance and/or exit channels.  While theory has demonstrated encouraging progress over the course of those many decades in describing separate, specific processes, one can imagine that for each separate symmetry (i.e. angular momentum and total parity, $J^\pi$) of this most fundamental neutral molecule, there exists a scattering matrix ${\cal S}(E)$ that has {\emph all} energetically open channels included.

Direct solution of the four-body problem with pairwise interactions  remains extremely challenging, of course.  But in the present case where the two nuclei are 3 orders of magnitude heavier than the two electrons, the Born-Oppenheimer approximation is an excellent starting point.  For the description of molecular vibrational levels lying in discrete potential curves, this is well-developed and high accuracy is routinely achieved.  But a fundamental difficulty arises when one attempts to apply the Born-Oppenheimer strategy to any process that involves the electronic continuum.  An example is an inelastic electron collision with the molecular cation that can excite or de-excite rovibrational levels or even lead to dissociation following electron capture in the process called dissociative recombination.\cite{Bates, guberman1991generation, LarssonOrel,kokoouline2011breaking,SchneiderH2HD,EDFT_HeHplus_2020,hornquist2024dissociative} 

When describing the vibrational motion in a discrete Born-Oppenheimer potential curve, one takes the energy eigenvalues $U_n(R)$ of the fixed-$R$ Hamiltonian at all $R$ values, and this serves as the potential energy in the vibrational Schr\"odinger equation.  A fundamental question arises, however, when describing an electron belonging to the continuous spectrum that collides with a molecular ion, as in the e$-H_2^+$ system.  One can compute either the body-frame (clamped-nucleus) scattering matrix, or equivalently the body-frame reaction matrix $K_{jj'}({\cal E},R)$, at all energies ${\cal E}$ and at all $R$ values, but one must convert, or ``frame-transform'' that information into a laboratory-frame scattering matrix or reaction matrix $K_{vj,v'j'}(E)$, which must depend only on the total laboratory frame energy $E$. A key question addressed in different ways over the years is the following:  At a given total energy $E$, which body-frame energies ${\cal E}$ (or more explicitly, ${\cal E}(R)$) contribute to the scattering process, and in what manner does $K_{jj'}({\cal E},R)$ map onto the desired $K_{vj,v'j'}(E)$? 

In a case such as the low energy {\it ungerade} states of H$_2$, where the neutral Rydberg states attached only to the $1s\sigma$ ionic ground state potential curve $V_{1s\sigma}^+(R)$ are 
relevant, an effective solution has been developed, starting with Ref.\cite{Gao_Greene_PRAR_1990} and refined with extensive tests in Ref.\cite{Backprop_FT_2020}, based on the following argument:  out of all of the possible trajectories in the ${\cal E},R$ plane that could be utilized in the calculation of vibrational frame transformation radial integrals such as $S_{vj,v'j'}(E)=\langle v,j | S_{j,j'}({\cal E}(R),R) |v',j' \rangle$, the one where the Born-Oppenheimer approximation is best will simply be the one where the nuclei feel {\it no  additional force}
during the short time when the electron is at short range (in the reaction zone); that choice amounts mathematically to choosing ${\cal E}(R)$ in the above matrix element to equal a constant shift from the ionic potential curve determined by the incident (or right-hand side) electron channel energy, namely ${\cal E}(R) = V_{1s\sigma}(R)+\epsilon_{v'}$, where $\epsilon_{v'}\equiv E-E_{v'}$ is the electron energy measured from the vibrational energy $E_{v'}$ of the positive molecular ion. Detailed evidence that this energy-dependent frame transformation strategy is effective in describing the challenging dissociative recombination (DR) process in the $H_2$ {\it ungerade} symmetry is presented in Ref.\cite{Backprop_FT_2020}.

Despite the demonstrated success of that strategy for treating systems with a single dominantly contributing ionic potential curve, a second frequently encountered class of problems involving electron scattering from a molecular ion remains unsolved in the framework of the energy-dependent frame transformation:  namely, situations where the ion has two or more potential energy curves that have Rydberg and scattering states attached to them; as is well-known, the $H_2$ {\it gerade} symmetry is the prototype of such systems.  For the gerade symmetry, there is at small $R$ the lowest doubly excited state $2p\sigma^2$, which is above the $1s\sigma$ ionic state and thus autoionizing for $R\lesssim 2.6 $ a.u. But it crashes down and becomes bound at $R \gtrsim 2.6$, causing a family of avoided crossings that create double-minimum structures in the {\it gerade} $H_2$  bound state potential energy curves such as the $EF$ and $GK$ states.  The strategy that was successful for the {\it ungerade} states is now ambiguous, because the two ionic potential curves are not parallel, and a choice of the $U(R)$ to use in frame transformation integrals is ambiguous.  The main point of the present study is to present a resolution of this ambiguity, a strategy that we demonstrate can successfully describe processes such as the dissociative recombination that occurs in the {\it gerade symmetry} of electron-$H_2^+$ collisions.  This initial benchmark study is carried out for an $H_2$-like model for the gerade symmetry, a minor revision of the two-dimensional model presented in \cite{EPJD_H2_Gerade_2022}, which has the advantage that it is exactly solvable (e.g. via the 2D $R$-matrix method \cite{Curik_HG_2DRmat_2018}) and can therefore provide a stringent test of the energy-dependent vibrational frame transformation theory developed here.

The present study connects naturally with many of the motivations that have propelled Christian Jungen and his collaborators in their quest to apply and extend the MQDT description of molecular hydrogen to make it as complete and comprehensive as possible.  Their work has had tremendous success in its applications that have focused primarily on ionization and dissociation processes that are probed in photon absorption.  For an impressive example, see \cite{SommavillaJungen2016} and references therein.  In their treatments of the H$_2$ $np$ {\it ungerade} singlet states, the Jungen collaboration has taken state-of-the-art calculations of H$_2$ potential energy curves $U_n(R)$, and converting them to quantum defect functions, $\mu_n(R)$, and then applying a rovibrational frame transformation that converts the information into photoionization and photodissociation spectra.  Despite the tremendous successes of that strategy, many molecular species and even some of the symmetries in H$_2$ and its isotopologues require additional theoretical and conceptual improvements.  For instance, in most molecules, the molecular ion possesses multiple low energy potential curves, which means that it becomes ambiguous to convert an accurate potential energy curve of the neutral molecule unambiguously into quantum defect or analogous phase information.  In such a situation one must presumably rely on direct {\it ab initio} computation of the multichannel body-frame scattering information in the form discussed above, i.e. $K_{jj'}({\cal E},R)$, at all energies ${\cal E}$ and at all $R$ values. Jungen and Ross\cite{Jungen_Ross_1997}  have also developed a novel approach that extends the initially-determined electron-ion scattering matrix to also include dissociation channels, allowing them to predict scattering processes involving molecular dissociation in the initial or final state.  So far their treatment appears to have had limited applicability to systems with only one dissociative potential curve per symmetry (e.g. $\Sigma$ and $\Pi$ in H$_2$).  Another goal of the present study is a reformulation of the Jungen-Ross treatment of ionization-dissociation channel coupling that can include multiple dissociation channels per molecular symmetry.  This treatment is also benchmarked against our essentially exact $R$-matrix solution of the 2D model of $^1\Sigma_g$ H$_2$, applied to vibrational excitation and dissociative recombination.

\section{Description of the H$_2$ gerade system}
\subsection{An improved two-dimensional model}
\begin{figure}[ht]
    \centering
    \includegraphics[width=1.0\linewidth]{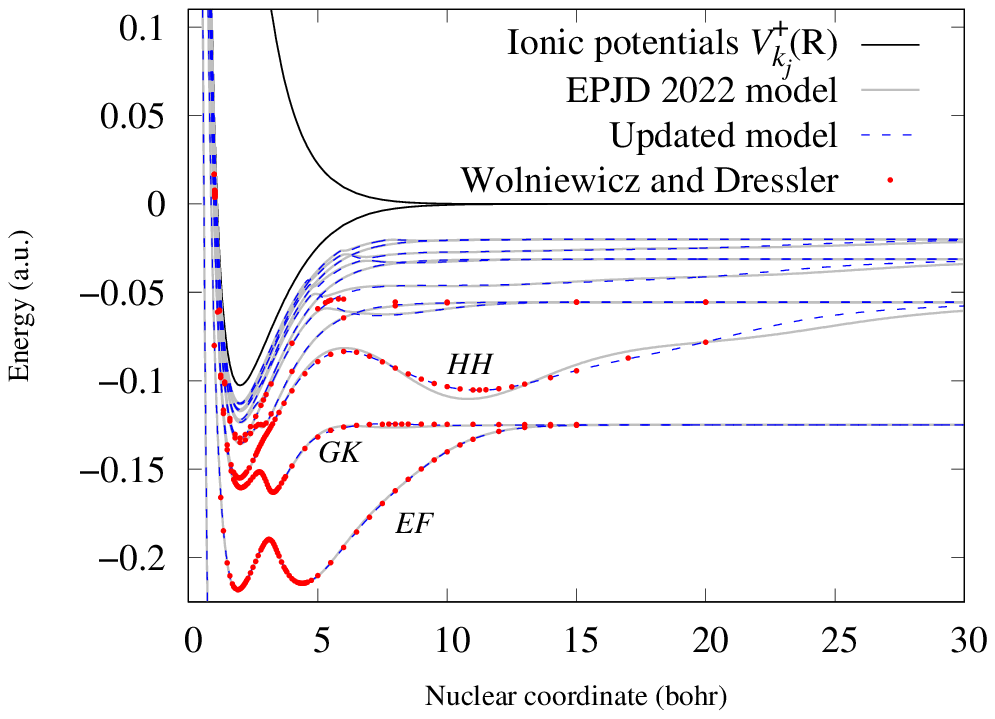}
    \caption{Comparison between neutral H$_2$ potential curves obtained using our updated model (blue dashed lines) versus our EPJD 2022 \cite{EPJD_H2_Gerade_2022} model (grey lines), both made to reproduce the data of Wolniewicz and Dressler \cite{Wolniewicz_Dressler_JCP_1985} $EF$, $GK$ and $H\bar{H}$ curves (red points). The image also contains the Wolniewicz and Dressler $O$ and $P$ curves (also red points) as well as the H$_2^+$ ion potentials for reference (black lines).}
    \label{fig-BOcurves}
\end{figure}

For this study, we will analyze the $^1\Sigma_g$ symmetry of H$_2$ which is a good example of a system with many interacting pathways for ionization and dissociation including direct and indirect dissociative recombination. To describe it a flexible model is introduced that can be solved two different ways: {\it (i)} by using our presented MQDT frame transformation technique;  or {\it (ii)} by employing a 2D R-matrix approach \cite{Curik_HG_2DRmat_2018} that avoids all the physically-motivated approximations and can be solved to essentially arbitrary precision, making it an ideal tool for method development and benchmarking. At the same time, we stress that this model closely resembles many aspects of this symmetry of the H$_2$ molecule. 
The model consists of a 2-dimensional Hamiltonian coupling multiple electronic angular momenta $l$ and two different ionic potential curves $V^+_k(R)$ and their corresponding eigenfunctions.  The Hilbert space kets $|j \rangle$ that incorporate those two degrees of freedom will be denoted with a collective index $j \equiv \{k,l\}$.  In MQDT language, these are designed to represent the three singlet channels of H$_2$, namely $|j=0 \rangle \equiv 1s\sigma \epsilon s \sigma  $ has $k=0$ and $l=0$, while $|j=1 \rangle \equiv 2p\sigma \epsilon s \sigma  $ has $k=1$ and $l=1$, and finally $|j=2 \rangle \equiv 1s\sigma \epsilon d \sigma  $ has $k=0$ and $l=2$.  In this MQDT-inspired notation, the use of $\epsilon$ to represent the scattering electron kinetic energy at infinity implies that the entire radial Hilbert space of the outermost electron is included, both the bound portion at $\epsilon<0$ and the continuum electron states at $\epsilon>0$ in each channel.  The matrix Hamiltonian for this model problem has the following structure:
\begin{equation}
\label{eq-Hamiltonian_2D}
\begin{split}
    \bra{j} \hat{H}(R,r) \ket{j'}= & \bra{j} \hat{H}^{\text{n}}(R) + \hat{H}^{\text{e}}(r) + \hat{V}^{\text{int}}(R,r) \ket{j'} \\
    = & \left[ H_j^{\text{n}}(R) + H_j^{\text{e}}(r)\right] \delta_{jj'} + V^{\text{int}}_{jj'}(R,r)\; ,
\end{split}
\end{equation}
where $R$ is the nuclear separation, $r$ is the distance of the outer electron from the molecule's center of charge and $j\in\{0,1,2\}$ is the collective channel index. The symbol $\delta_{jj'}\equiv \delta_{kk'}\delta_{ll'}$ denotes the composite Kronecker delta and the diagonal terms are the electronic and nuclear Hamiltonians
\begin{equation}
\label{eq-Hamiltonian_elnuc}
\begin{gathered}
    H_j^{\mathrm{e}}(r) = - \frac{1}{2}\frac{\partial^2}{\partial r^2} - \frac{1}{r} + \frac{l_j(l_j+1)}{2r^2}\;, \\
    H_j^{\mathrm{n}}(R) =  - \frac{1}{2M}\frac{\partial^2}{\partial R^2} + V^+_{k_j}(R)\; ,
\end{gathered}
\end{equation}
where $M$ is the nuclear reduced mass and $V^+_{k_j}(R)$ are the ionic potential curves (taken from e.g. Madsen and Peek \cite{Peek_Madsen_1971}) of the $1s\sigma_g$ state for $j\in\{0,2\}$ and the $2p\sigma_u$ state for $j=1$. The final term of (\ref{eq-Hamiltonian_2D}) is the electron-nuclear interaction potential which couples all of the degrees of freedom at short ranges. It is based on the model interaction potential from our previous study \cite{EPJD_H2_Gerade_2022} with some small alterations introduced in order to more closely reproduce the relevant H$_2$ gerade potential curves\cite{Wolniewicz_Dressler_JCP_1985}. Its general form is
\begin{equation}
\label{eq-model_potential}
    V^{\text{int}}_{jj'}(R,r) = e^{-r^2/\omega^2}\,
    \sum_{i=1}^3 a_i e^{-\left( \frac{R-b_i}{c_i}\right)^2}
\end{equation}
for any combination of $j,j'$.
The width of the electronic Gaussian is always fixed at $\omega=2\text{ bohr}$ and the constants $a,b,c$ are given in Table \ref{tab-model_V}. The improvement in accuracy over the previous version of the model can be seen in Fig.~\ref{fig-BOcurves}.
Note that the symmetric potential coupling matrix is non-negligible only at short range in both coordinates, $r<r_b, R<R_b$.
\begin{table}[ht]
    \setlength{\tabcolsep}{6pt}
    \centering
    \begin{tabular}{crrrrr}
        \hline
         & $V_{00}$\pp & $V_{11}$\pp & $V_{22}$\pp & $V_{01}$\pp & $V_{12}$\pp \\
        \hline
        $a_{1}$ & 0.31858 &     -0.67302  &     -0.81074  & 0.22626 & 0.50704 \\
        $b_{1}$ & 3.93449 & \phm 3.33607 & \phm 8.64348  & 2.41245 & 2.72802 \\
        $c_{1}$ & 1.64285 & \phm 2.92052 & \phm 6.05749  & 1.41711 & 1.69415 \\
        $a_{2}$ & 0.13262 &     -0.34016 &     -1.69468  & 0 \pp    & 0 \pp   \\
        $b_{2}$ & 5.83366 & \phm 7.19246 & \phm 18.16842 & - \pp    & - \pp   \\
        $c_{2}$ & 2.46164 & \phm 3.52307 & \phm 12.87950 & - \pp    & - \pp   \\
        $a_{3}$ & 0.15261 & \phm 0 \pp   &     -0.21033  & 0  \pp   & 0 \pp   \\
        $b_{3}$ & 2.47634 &  - \pp       & \phm 5.53487  & -  \pp   & - \pp   \\
        $c_{3}$ & 0.97500 &  - \pp       & \phm 3.22183  & -  \pp   & - \pp    \\
        \hline
    \end{tabular}
    \caption{Parameters of the 2D model potential. The direct $s$--$d$ coupling is neglected in the present model. Therefore, the column $V_{02}$ is omitted.}
    \label{tab-model_V}
\end{table}

The benchmark solution of the Schr\"odinger equation through the 2D R-matrix method has been adequately described in our previous study \cite{EPJD_H2_Gerade_2022}, the only difference here being our slightly modified and more realistic model Hamiltonian for the $^1\Sigma_g$ symmetry that is described above.
We regard that benchmark 2D R-matrix solution as ``essentially exact'', and now the rest of this section describes the various quantities necessary for a multichannel quantum defect theory description of the system.

\subsection{MQDT description of the system}
To characterize the system with MQDT methods, our calculations of the nontrivial electron-ion interactions are carried out within a box defined by $r \leq r_0, R \leq R_0$ combined with matching to asymptotic solutions outside. The boundary $r_0$ needs to be large enough to encompass all electron-nuclear interaction ($r_0 \geq r_b$) but also small enough to accommodate the Born-Oppenheimer approximation. The boundary $R_0$ needn't be larger than $R_b$, but it must be large enough for the off-diagonal terms of $V^{\text{int}}_{jj'}(R,r)$ to be negligible at $R>R_0$. This means that nuclear fragmentation channels are decoupled beyond $R_0$.  Our treatment is designed so far to treat the low energy range where three-body fragmentation cannot occur, i.e. in any final channel of the physical $S$ matrix, either $r$ or $R$ can escape to infinity, but not both.

The first important quantity required for an MQDT description are the nuclear vibrational eigenfunctions in the ionic potential curves. They satisfy the radial 1D Schr\"odinger equation in the respective potentials
\begin{equation}
\label{eq-vibrational_chi}
    \hat{H}^{\text{n}}_j(R) \; \chi_{v,j}(R) \ket{j} = E_{v,j} \; \chi_{v,j}(R) \ket{j},
\end{equation}
with some outer boundary condition at $R=R_0$ that is defined  later.
$E_{v,j}$ are the vibrational channel thresholds and we combine the $\{v,j\}\equiv \{v,k_j,l_j\}$ into a single composite index $i$ describing the full ionic vibrational basis.  Because most of our effort is devoted to scattering calculations, for which the energy is in the continuum, energy appears in multiple different contexts, and it is helpful to use a notation that distinguishes the different types of energies.  In particular, the variable $E$ is always the total physical energy of the system, e.g. in the final physically relevant scattering matrix, referred to the asymptote H(1s)+p +e at $R\rightarrow \infty$.  Instead, the energy is referred to as ${\cal E}$ in the framework of the Born-Oppenheimer potential curves, e.g. at fixed $R$, and with the same choice of zero. Other energies that arise in the description below are the electronic channel energies, e.g. $\epsilon_i = E-E_i$ is the asymptotic electron kinetic energy when the H$_2^+$ ion is left in a particular vibrational energy level $E_i$ of a specific potential curve with index $k_i$.  The electronic channel energies at fixed internuclear distance will instead be designated by the variable
$\varepsilon_j$ = ${\cal E}-V_j^+(R)$.

Another one-dimensional Schr\"odinger equation that plays a key role is the Born-Oppenheimer electronic energy eigenvalue equation for the neutral molecule; this is in fact 3 coupled differential equations in $r$ for each fixed value of $R$:
\begin{equation}
\label{eq-BO_functions}
    \hat{H}^{\text{BO}}(R;r) \ket{\phi_{d}^{\text{BO}}(R;r)} = U_d(R) \ket{\phi_{d}^{\text{BO}}(R;r)}.
\end{equation}
Here the index $d$ labels the eigenfunctions and potential curves, some of which can serve as dissociative channel potentials, as is seen below. In the preceding equation, $H^{\text{BO}}$ is the full Hamiltonian of the system evaluated at some fixed $R$ but with the nuclear kinetic term omitted.
Evidently, from the structure of the full Hamiltonian, one sees that the ket $\ket{\phi_{d}^{\text{BO}}(R;r)}$ contains, in general, a non-trivial combination of H$_2$ channel kets $\ket{j}$ multiplied by radial functions. We can also rewrite
\begin{align}
    \nonumber
   \ket{ \phi_{d}^{\text{BO}}(R;r)} &= \sum_{j'} \phi_{d,j'}^{\text{BO}}(R;r) \ket{j'} \;, \\
    &\downarrow \nonumber \\
    \bra{j} \hat{H}^{\text{BO}}(R;r) \ket{\phi_{d}^{\text{BO}}(R;r)} &= U_d(R) \langle j \ket{\phi_{d}^{\text{BO}}(R;r)}\,, \nonumber \\
    &\downarrow \nonumber \\
    \left[ H_j^{\text{e}}(r) + V^+_{k_j}(R) \right] \phi_{d,j}^{\text{BO}}(R;r) &+ \sum_{j'} V^{\text{int}}_{jj'}(R,r) \phi_{d,j'}^{\text{BO}}(R;r) \nonumber \\
    \label{eq-BO_functions_rewrite}
    &= U_d(R) \phi_{d,j}^{\text{BO}}(R;r)
\end{align}
for any fixed $R$. This equation will be solved in various regions, with various boundary conditions applied.

If we impose vanishing boundary conditions at large $r>r_0$, then the fixed-nuclei eigenvalues obtained are ${\cal E}=U_d(R)$, i.e these are the usual Born-Oppenheimer potential curves plotted in Fig. \ref{fig-BOcurves}.
For $R>R_0$ nonadiabatic coupling is neglected, and the curves $U_d(R)$ represent the paths that the neutral system can take to dissociate. Thus we compute these even outside the $R_0$ box so that we can calculate Milne-type (described in section IV. B of Ref. \cite{Greene_Rau_Fano_1982}) or WKB type radial solutions following these curves.
The quantity $\phi_{d}^{\text{BO}}(R;r)$ itself is mostly important at the dissociative surface $R=R_0$ and for large values of $r>r_0$, as will be discussed in section \ref{subsection-extract_boundary}.

The final $R$-fixed MQDT quantity needed is the Born-Oppenheimer short-range $K$ matrix, which is computed in the manner described in sections 3.3 and 3.4 in our 2D model study \cite{EPJD_H2_Gerade_2022}. In short, we solve Eq. (\ref{eq-BO_functions}) inside the box $r \leq r_0, R \leq R_0$ for the symmetrized Hamiltonian
\begin{equation}
    \bar{H}^{\text{BO}}(R;r) = \hat{H}^{\text{BO}}(R;r)+\frac{1}{2}\delta(r-r_0)\frac{\partial}{\partial r}\,,
\end{equation}
while imposing no boundary condition at $r=r_0$.
The eigensolutions are plugged into the Wigner-Eisenbud expansion \cite{Wigner_Eisenbud_PR_1947} to obtain a fixed-nuclei, ${\cal E}$-dependent $R$ matrix that is non-diagonal in the index $j$.  The $R$ matrix is then transformed into the Born-Oppenheimer short-range matrix $K^{\text{BO}}_{jj'}({\cal E},R)$ by matching to electronic Coulomb functions at $r=r_0$.
Considering the strongly closed channels that will be a part of our method, it is advisable to switch from the standard energy-normalized pair of coulomb functions $\{f,g\}$ ($\{\sqrt{2}s,-\sqrt{2}c\}$ in Seaton's work \cite{Seaton_2002})to an alternate pair with smoother behavior at negative energies. We denote them $\{f^0,g^0\}$ (equal to $\{f,-h\}$ from Seaton \cite{Seaton_2002}). They differ from the energy-normalized pair by the relation $\{f^0,g^0\}=\{f/q,g \cdot q\}$, where $q$ denotes the square root of the energy- and angular-momentum-dependent factor $B(\epsilon,l)$ from Seaton's work \cite{Seaton_2002}. The resulting quantum defect is typically denoted as $\eta$. We can diagonalize the reaction matrix $K^{\text{BO}}_{jj'}({\cal E},R)$ to compute the quantum defect and related sine and cosine matrices $\mathcal{S},\mathcal{C}$ as
\begin{align}
    \nonumber
    K^{\text{BO}}_{jj'} &= \sum_\alpha U_{j\alpha} \tan(\pi \eta_\alpha) \left(U^{\text T}\right)_{\alpha j'}, \\
    \nonumber
    \eta_{jj'} &= \sum_\alpha U_{j\alpha} \eta_\alpha \left(U^{\text T}\right)_{\alpha j'}, \\
    \nonumber
    \mathcal{S}_{jj'} &= \sum_\alpha U_{j\alpha} \sin(\pi \eta_\alpha) \left(U^{\text T}\right)_{\alpha j'}, \\
    \label{eq-QDF_matrix}
    \mathcal{C}_{jj'} &= \sum_\alpha U_{j\alpha} \cos(\pi \eta_\alpha) \left(U^{\text T}\right)_{\alpha j'},
\end{align}
while picking continuous branches of the eigendefect $\eta({\cal E},R) = \arctan(\tan(\pi\eta({\cal E},R))/\pi$ with respect to both energy and coordinate.
The matrices $\mathcal{S,C}$ do not have poles like the reaction matrix $K^{\text{BO}}$ and they are much easier to interpolate from stored values. We need to compute and store these matrices over a wide array of energies and values $0\leq R \leq R_0$.
It is useful to remember that the matrices in Eq. (\ref{eq-QDF_matrix}) evaluated at Born-Oppenheimer energy ${\cal E}$ are used to represent the fixed-nuclei channel functions at $r>r_0$ in terms of Coulomb functions evaluated at energies $\varepsilon_j={\cal E}-V^+_{k_j}(R)$.

\section{\label{sec-EDFT}Energy-dependent frame transformation for gerade H$_2$}

The following outlines our frame-transformation formulation, and then motivates the need to explicitly incorporate the body-frame quantum defect energy dependence, and then segues into our implementation of the modified Jungen-Ross method of Section \ref{sec-basic_ideas} for incorporating molecular dissociation channels.
The energy-dependent frame transformation approach is based on the outline of Ref.\cite{Gao_Greene_PRAR_1990} and is an expanded version of our previous ``simplified" energy-dependent frame transformation first tested for the ungerade symmetry of H$_2^+$ \cite{Backprop_FT_2020} and then implemented for HeH$^+$ \cite{EDFT_HeHplus_2020}. The expansion comes from the presence of three coupled electronic partial waves, which results in the extension of the quantum defect (and all related quantities) into a quantum defect matrix as seen in Eq. (\ref{eq-QDF_matrix}).

\begin{figure}[tbh]
    \centering
    \includegraphics[width=1.0\linewidth]{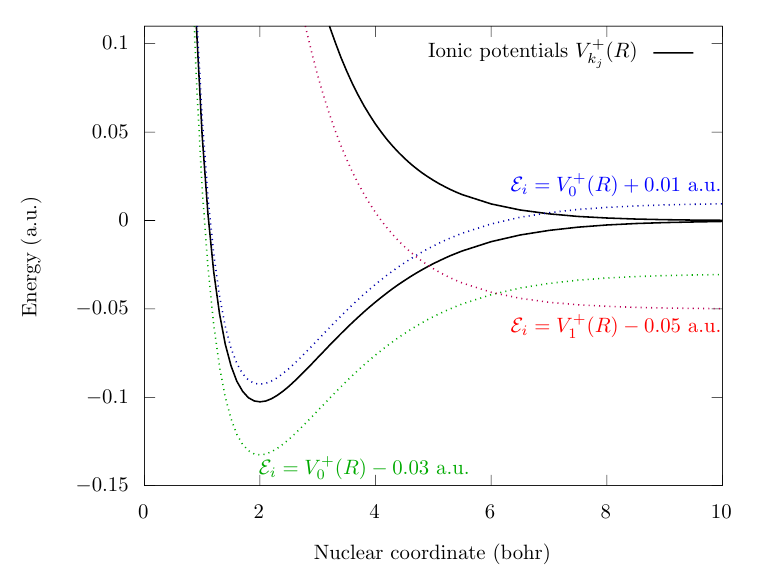}
    \caption{Example of body-frame energy curves ${\cal E}_{\epsilon,k}(R) = \epsilon+V_{k}^+(R)$ parallel to the ionic potentials. The blue dashed line is an example of an open channel with $\epsilon=0.01$~a.u. and $k=0$. The green dashed line represents a closed channel with $\epsilon=-0.03$~a.u., $k=0$ and the red dashed line represents a closed channel with $\epsilon=-0.05$~a.u., $k=1$.}
    \label{fig-E_parallel}
\end{figure}

For the total energy $E$ of the system in the laboratory frame we define the ionization channel energies of an electron in the field of a quantized ionic vibrational level as $\epsilon_i=E-E_{v_i,j_i}$ ; moreover, body frame energy curves {\it parallel to the $k$-th ionic potential}, as shown in Fig. \ref{fig-E_parallel}, will be of particular interest in the following, e.g., ${\cal E}_{\epsilon,k}(R) = \epsilon+V_{k}^+(R)$. Such parallel choices, out of the continuously infinite set of possible values of the above matrices ${\mathcal S}({\cal E},R),
{\mathcal C}({\cal E},R)$ that were initially computed for {\it all} ${\cal E},R$, play a 
central role in this formulation of the simplified EDFT. Specifically, the theory takes the QDT information of Eq. (\ref{eq-QDF_matrix}) that was computed above, and constructs a set of Born-Oppenheimer solutions that obey the Schr\"odinger equation at small electronic distances and at total energy $E$, and which are characterized by matrices $\mathcal{S,C}$ on the electronic surface $r=r_0$.  The $i$-th such independent solution at this energy $E$ has the following form on the electronic surface of the reaction volume:
\onecolumngrid
\begin{equation}
\label{eq-EDFT_psi_BO}
    \Psi^{\text{BO}}_i(E,r;R) = {\cal N}_i(E,R)\chi_i(R) \sum_{j'} \ket{j'}
     \bigg( f^0(l_{j'},r) \mathcal{C}_{j' j_i}({\cal E}_{\epsilon_i,k_i}(R),R) - g^0(l_{j'},r) \mathcal{S}_{j' j_i}({\cal E}_{\epsilon_i,k_i}(R),R) \bigg) .
\end{equation}
\twocolumngrid
Note that no electron energies have been specified in the Coulomb functions, because they will eventually be understood to be the ``back-propagated'' solutions at $r=0$, in the spirit of Ref.\cite{Backprop_FT_2020}, whereby they have negligible energy dependence.  Crucially, the curves in the ${\cal E},R$ plane ${\cal E}_{\epsilon_i,k_i}(R)$ have intentionally been chosen to lie parallel to $V^+_{k_i}(R)$ 
according to the logic of Ref.\cite{Gao_Greene_PRAR_1990}, i.e., for this choice, in the $i$-th linearly-independent solution to the Schr\"odinger equation, the nuclei experience {\it no additional force} (at least for the part of the solution in channel $i$) in $R$ during the brief period that the electron spends inside the reaction volume. This choice also allows us to use the ionic vibrational channel functions $\chi_i(R)$ as the Born-Oppenheimer radial wavefunction in the solution. The symbol ${\cal N}_i(E,R)$ denotes a normalization factor.  These solutions are matched on the electronic surface to general outer region ($r\geq r_0$) solutions (and derivatives), giving
\begin{multline}
\label{eq-EDFT_psi_out}
    \Psi^{\text{out}}_i(E,r,R) = \sum_{i'} \ket{j_{i'}}\chi_{i'}(R)  \bigg(f^0(\epsilon_{i'},l_{i'},r) \mathcal{C}^{\text{FT}}_{i' i}(E) \\
    - g^0(\epsilon_{i'},l_{i'},r) \mathcal{S}^{\text{FT}}_{i' i}(E) \bigg),
\end{multline}
with unknown $\mathcal{S}^{\text{FT}},\mathcal{C}^{\text{FT}}$.
Matching Eq. (\ref{eq-EDFT_psi_BO}) and (\ref{eq-EDFT_psi_out}), and projecting onto $\bra{j_{i''}}\chi_{i''}(R)$ results in more complicated forms of these matrices. The "simplified" energy-dependent frame transformation approach then assumes ${\cal N}_i(E,R)\approx 1$ (as in Ref.\cite{Gao_Greene_JCP_1989}) and negligible energy dependence of Coulomb function Wronskians, reasonable at small enough electronic box sizes, to yield (abbreviating ${\cal E}_{\epsilon_i,k_i}(R) \equiv  {\cal E}_i $):
\begin{align}
    \nonumber
    \mathcal{S}^{\text{FT}}_{ii'}(E) &= \int_{0}^{R_0}  \chi_{i}(R) \mathcal{S}_{j_{i}j_{i'}}(\mathcal{E}_{i'},R) \chi_{i'}(R)  \text{d}R, \\
    \label{eq-EDFT_match}
    \mathcal{C}^{\text{FT}}_{ii'}(E) &= \int_{0}^{R_0}  \chi_{i}(R) \mathcal{C}_{j_{i}j_{i'}}(\mathcal{E}_{i'},R) \chi_{i'}(R) \text{d}R.
\end{align}
These can be combined into the frame-transformed reaction matrix $K^{\text{FT}} = \mathcal{S}^{\text{FT}} (\mathcal{C}^{\text{FT}})^{-1}$ by recasting Eq. (\ref{eq-EDFT_psi_out}) into
\begin{multline}
\label{eq-EDFT_psi_Kform}
    \Psi^{(K)}_i(E,r,R) = \sum_{i'} \chi_{i'}(R) \ket{j_{i'}} \bigg( f^0_{i'}(\epsilon_{i'},l_{i'},r) \delta_{i' i} \\
    - g^0_{i'}(\epsilon_{i'},l_{i'},r) K^{\text{FT}}_{i' i}(E) \bigg).
\end{multline}

Eqs.(\ref{eq-EDFT_psi_out}), (\ref{eq-EDFT_match}), and (\ref{eq-EDFT_psi_Kform}) constitute our generalization of the energy-dependent frame transformation (EDFT) that has previously been implemented only for MQDT systems with a single molecular ion potential curve.  Now, armed with this generalization, we can proceed to apply this to the more complex situation such as the gerade $^1\Sigma_g$ symmetry of H$_2$ which involves multiple ionic potential curves.

Observe that the unphysical reaction matrix couples both open and closed channels and due to the right-index energy dependence of Eq. (\ref{eq-EDFT_match}), it is not manifestly symmetric and we have found that best results are obtained if it is symmetrized manually at this point.  Here, as in other MQDT studies, when we use the term ``unphysical'', it simply means that the final boundary condition forcing the wavefunction to remain finite at infinite distance has not yet been imposed.  If there were no dissociation channels relevant to be included in the final scattering matrix, one could apply the usual MQDT closed-channel elimination formula to get the highly $E$-dependent physical electron-ion scattering matrix; but in many cases there are dissociation channels open which should also be included, some of which might be weakly-closed as well, and this will be addressed in the following Section.

For an energy-independent approach, one would follow this same procedure while neglecting the energy dependence in the QDT data of Eq. (\ref{eq-QDF_matrix}) and in the following steps, the frame-transformed ${\cal S},{\cal C}$ matrices would also be evaluated at just one fixed body-frame energy.

\section{Incorporation of the full energy dependence and a reformulation to include dissociation}

\begin{figure}[ht]
    \centering
    \includegraphics[width=1.0\linewidth]{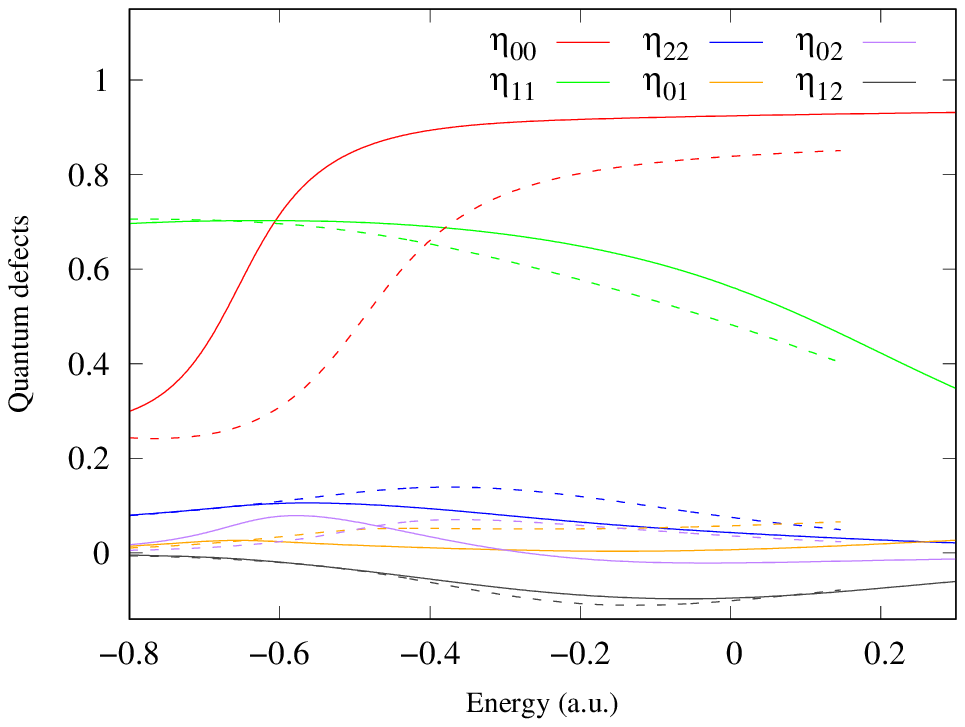}
    \caption{The energy dependence of the quantum defect matrix $\eta_{jj'}({\cal E},R)$ at two $R$ values. The full curves are evaluated at $R=2$ bohr and the dashed curves at $R=3$ bohr. In this graph, energy equal to zero has the same meaning as in Fig. \ref{fig-BOcurves}.}
    \label{fig-QDF_edep}
\end{figure}

Of the two relevant ionic potential curves the lower potential curve $V_{0}^+(R)$ supporting $s,d$-waves of the outermost electron contains the bound vibrational states of the molecule while the upper mostly repulsive $p$-wave curve $V_{1}^+(R)$ is responsible for the $2p\sigma^2$ \cite{RossJungen1994} (and higher) resonance, providing the direct DR pathways. The $2p\sigma$ potential curve of H$_2^+$ does support two long-range weakly-bound states, as has been discussed elsewhere, but those are largely irrelevant in the range of energies considered here.\cite{carbonell2003new,beyer2018hyperfine,singh2024adiabatic}. 
In order to describe wavefunctions following this dominant direct DR mechanism in the Franck-Condon region it is necessary to include highly excited $p$-wave vibrational channels in our solutions in Eqs. (\ref{eq-EDFT_psi_BO}) and (\ref{eq-EDFT_psi_out}). This results in evaluating the QDT information of Eq. (\ref{eq-QDF_matrix}) at some very strongly closed energies.  It is important that the $2p\sigma$ potential curve vibrational levels with an inner turning point can at least reach the Condon point, which is at $R \approx 2.0$ a.u.
The open ionization channels, on the other hand, require low lying bound states of the $s,d$-curve $V_0^+(R)$.
The frame transformations in Eq. (\ref{eq-EDFT_match}) at a fixed lab-frame energy follow curves ${\cal E}_{\epsilon_i,k_i}(R)$ parallel to the ionic potentials which are qualitatively different for the $p$-wave compared to the $s$- and $d$-waves. Fig. \ref{fig-QDF_edep} shows the energy dependence of $\eta_{jj'}$ at two $R$ values within the Franck-Condon region. Clearly, the integrals of Eq. (\ref{eq-EDFT_match}) will require very different input matrices depending on the channel indices $ii'$ and treating this problem with energy-independent body-frame quantum defect matrices would not be sensible.

Most FT applications to date that computed DR cross sections have utilized a vibrational basis with an outgoing-wave boundary condition such as Siegert pseudostates \cite{Tolstikhin_siegert_1998} or an exterior complex scaling basis \cite{McCurdy_Martin_JPB_2004}.
Here we adapt a different method in which all channel thresholds are strictly real, the scattering matrix is unitary, and which also allows us to differentiate between partial DR cross sections.

\subsection{\label{sec-basic_ideas}Basic ideas of the 1997 Jungen-Ross treatment}

Extensive theoretical work by Christian Jungen and his collaborators, especially Stephen Ross, has led in 1997 to an improved capability of MQDT to handle competing ionization and dissociation channels of the H$_2$ molecule \cite{Jungen_Ross_1997}.  In this section, we revisit their formulation and make some minor revisions and extensions of that treatment.

The key starting point begins with the success of the Jungen-Dill-Fano rovibrational frame transformation\cite{Fano1970, ChangFano, Jungen_Dill_JCP_1980, GreeneJungen1985adv} based on multichannel quantum defect theory (MQDT) to describe photoionization of H$_2$, especially in the vicinity of rotationally- and vibrationally-autoionizing Rydberg states attached to rovibrational thresholds $E_{v_j,N_j}$.  In the following discussion, we will omit reference to the nuclear rotational degrees of freedom $N_j$, so the vibrational thresholds will be labeled simply as $E_{i}$, with corresponding vibrational eigenstates $\ket{i} \equiv  \ket{j_{i}}\chi_i(R)$ of H$_2^+$.
Remarkably, in the approximation that the fixed-nuclei quantum defect function $\mu(R)$ for the relevant symmetry is independent of the body-frame energy $\epsilon$, the electron-ion scattering matrix $S_{ii'}$ and other related matrices of scattering theory (such as the reaction matrix $K_{ii'}$ can be immediately obtained through matrix elements of the quantum defect function in the ionic vibrational basis set, i.e.:
\begin{equation}
    S_{ii'}= \langle i | e^{2 i \pi \mu(R) } | {i}' \rangle .
\end{equation}
Once this equation is obtained, with channels $i$ both open and closed, the standard closed-channel elimination formulas of MQDT determine the physical scattering matrix in the open channels only, but including the many closed-channel Rydberg resonances, e.g.:
\begin{equation}
\label{eq-channel_elimination_S}
     S^{phys}_{ii'}= S_{oo}-S_{oc}(S_{cc}-e^{-2 i \pi \nu_c})^{-1} S_{co}\,,
\end{equation}
where $o,c$ denote blocks of open and closed channels and $\nu_c$ is a diagonal matrix with the effective quantum numbers in the closed vibrational channels, i.e. in a.u., $\nu_{i} = [2(E_{i}-E)]^{-1/2}$ with $E$ representing the desired total energy of the system in the laboratory frame and with $E_i$ the ionic energy level in the $i$-th channel.

This energy-{\it in}dependent treatment, after photoabsorption dipole matrix elements are included in a straightforward manner, has had many successes in describing the photoabsorption and photoionization processes of H$_2$, starting with the impressive article by Jungen and Dill in 1980 \cite{Jungen_Dill_JCP_1980}, and with subsequent important applications by Jungen and Raoult \cite{Raoult_Jungen_JCP_1981}, and an application to HD photoionization by Ref.\cite{Du_Greene_JCP_1986}.

The H$_2$ molecule, of course, can {\it dissociate} as an alternative fragmentation pathway, at all energies where ionization is energetically allowed, but the standard frame-transformed scattering matrix has only ionization channels present.  However, the successes of the theory in treating ionization processes including all rovibrational interactions suggested to Jungen that the wavefunctions obtained in this theory are very accurate and should contain information about the dissociation channels.  The problem Jungen set out to solve in the early 1980s was to overcome this limitation of the theory, so as to allow the extraction of dissociation information from those accurate wavefunctions.  Jungen's first treatment accomplished this partially, with some successes, beginning with his 1984 article \cite{Jungen1984prl}, and further elaborated in the 1985 review article  \cite{GreeneJungen1985adv}.  Later efforts to explore dissociative processes included an article by Gao {\it et al.} that was published in 1993 \cite{GaoJungenGreene1993}.  All of the articles mentioned in this paragraph suffered from limitations and some awkwardness which left all of them short from being a robust, readily implemented theoretical framework.

But a key breakthrough came in the aforementioned 1997 PRA article by Jungen and Ross (JR), which yields a way to directly build the larger scattering matrix that includes both dissociation channels $d$ and ionization channels $i$ in a consistent manner. The desired (unphysical) scattering matrix  (called unphysical because it still contains some weakly-closed channels to be eliminated later) is constructed in the JR method to have the following block structure:
\begin{equation}
\label{eq-Smat_expand}
    {S}=
    \begin{bmatrix}
        S_{ii} & S_{id} \\
        S_{di} & S_{dd} \\
    \end{bmatrix}.
\end{equation}
Here $S_{ii}$ is a block that contains the ionization-only scattering matrix of the type computed in a finite range $0 < R \le R_0$ with vanishing boundary conditions (normally) at $R_0$, which is more or less identical to the original Jungen-Dill formulation $S$-matrix with ionization channels only. 
The new elements of Eq. (\ref{eq-Smat_expand}) involving dissociation channels are obtained by carrying out a new calculation of the Jungen-Dill type,\cite{Jungen_Dill_JCP_1980} but with different boundary conditions on the vibrational wavefunctions, for instance vanishing $R$-derivative at $R_0$ (Jungen and Ross refer to this second set of solutions as having a different $R$-log-derivative set equal to $-b^{(x)}$, which equals zero for the vanishing derivative choice just mentioned.). Then a slightly tricky procedure was used by Jungen and Ross to extract those new elements  $S_{id},S_{di},S_{dd}$, which we will next summarize before making our own adjustments. For starters, it is convenient to shift from the $S$-matrix form of the Jungen-Dill solutions $\Psi^{(S)}$ to the $K$-matrix form :
\begin{equation}
\label{eq-EDFT_psi_Kform2}
    \Psi^{(K)}_{i'}=\sum_i \ket{j_{i}} \chi_i(R) {\bigg (} f_i(r) \delta_{i{i'}}-g_i(r) K_{i{i'}} {\bigg )}.
\end{equation}
If our calculation is to include $N$ ionization channels, then the $ \Psi^{(K)}_{i'} $ represent  $N$ independent solutions, each of which can be viewed as one column of the radial solution matrix in the large parentheses. The first step is to separate those $N$ channels into the lowest $N_P$ open and ``weakly-closed'' thresholds, each of whose vibrational wavefunctions and their radial derivatives are totally negligible at $R_0$, and the remaining $N_Q=N-N_P$  ``strongly-closed'' channels.  It is the electronically strongly-closed portion of the Hilbert space in our calculation that is mostly responsible for representing the dissociating state at $R<R_0$, because the dissociating state(s) have low principal quantum number(s) of order $n \lesssim 2-3$.

The next step of the calculation is to impose exponential decay on all laboratory frame solutions in the $Q$-subspace.  This is achieved through the usual equation of MQDT
\begin{equation}
\label{eq-channel_elimination_K_basic2}
    {\mathcal{K}_{PP}}= K_{PP}-K_{PQ}(K_{QQ}+\tan{\beta_Q})^{-1} K_{QP},
\end{equation}
where $\tan{\beta_Q}=\tan{\pi(\nu_Q-l)}$ is a diagonal matrix needed to eliminate the exponentially growing terms.  Note that while we will have eliminated the exponentially growing terms in the $Q$-channel space after making this step, we will still need the coefficients of the exponentially decaying ``energy normalized'' Whittaker functions of $r$ in the $Q$-channels.  

The following sets up the required algebra, in terms of the radial solution matrices ${\underline F(r)}$ in the large parentheses of Eq. (\ref{eq-EDFT_psi_Kform2}), each of whose columns represents an independent solution of the Schr\"odinger equation, which allows us to determine the coefficients of the exponentially decaying solutions at $r \rightarrow \infty$:
\begin{equation}
\label{eq-solution_F}
    {\underline F(r)}\equiv  \left( \begin{array}{cc}
      f_P-g_P K_{PP} & -g_P K_{PQ} \\
      -g_Q K_{QP} & f_Q-g_Q K_{QQ} \\
  \end{array} \right)
\end{equation}
and the channel elimination step acts on this solution matrix from the right with a rectangular $N \times N_P$ matrix that eliminates exponential growth in all of the $Q$ channels, and produces $N_P$ independent solutions. To carry this out, first write the $Q$ channel Coulomb functions $f,g$ in terms of energy normalized rising/falling Whittaker solutions that are positive at $r\rightarrow \infty$, i.e. ${\bar W}_Q,W_Q$ respectively:
\begin{alignat}{2}
    \nonumber
    f_Q &= &{\bar W}_Q& \sin{\beta_Q} - W_Q \cos{\beta_Q}, \\
    \label{eq-Coulomb_whittaker}
    g_Q &= -&{\bar W}_Q& \cos{\beta_Q} - W_Q \sin{\beta_Q},
\end{alignat}
We can see that the matrix needed to right multiply ${\underline F(r)}$ that will eliminate the divergent ${\bar W}_Q$ terms and leave the $P$ space terms in standard $K$-matrix format must be
\begin{equation}
\label{eq-channel_elimination_A}
    \left( 
    \begin{array}{c}
    {\bf 1}_{PP} \\ 
    A_{QP}%
    \end{array}%
    \right) =\left( 
    \begin{array}{c}
    {\bf 1}_{PP} \\ 
    -(K_{QQ}+(\tan \beta_{Q}))^{-1}K_{QP}
    \end{array}%
    \right) 
\end{equation}
giving for our new solution matrix the following:
\begin{align}
    \nonumber
    {\mathcal F(r)} &= \left( 
   \begin{array}{cc}
    f_{P}-g_{P}K_{PP} & -g_{P}K_{PQ} \\ 
    -g_{Q}K_{QP} & f_{Q}-g_{Q}K_{QQ}%
    \end{array}%
    \right) \left( 
    \begin{array}{c}
    {\bf 1}_{PP} \\ 
    A_{QP}%
    \end{array}%
    \right) \\
    \label{eq-channel_elimination_F}
    &= \left( 
    \begin{array}{c}
    f_{P}(r)-g_{P}(r)\mathcal{K}_{PP} \\ 
    W_{Q}(r)Z_{QP}%
    \end{array}%
    \right) 
\end{align}
with 
${\mathcal{K}_{PP}}$ in the form of Eq. (\ref{eq-channel_elimination_K_basic2}), and the decaying Whittaker function coefficients
\begin{multline}
\label{eq-Zmat}
    Z_{QP}= \sin{\beta_Q} K_{QP}+(\cos{\beta_Q}-\sin{\beta_Q} K_{QQ}) \\
     \times (K_{QQ}+\tan{\beta_Q})^{-1}K_{QP}.
\end{multline}
Again, these solutions to the Schr\"odinger equation in the region $R<R_0$  have the boundary condition $\Psi\rightarrow 0$ at $R\rightarrow R_0$ applied.
It is necessary to note here, that when working with the alternate Coulomb function pair $\{f^0,g^0\}=\{f/q,g \cdot q\}$, essentially replacing Eq. (\ref{eq-EDFT_psi_Kform2}) with Eq. (\ref{eq-EDFT_psi_Kform}), then the previous algebra is slightly altered, yielding
\begin{align}
    \nonumber
    \mathcal{K}_{PP} &= q_P \left[ K^{\eta}_{PP}-K^{\eta}_{PQ}(K^{\eta}_{QQ}+\tan{\beta_Q}/q^2_Q)^{-1} K^{\eta}_{QP} \right] q_P, \\
    \nonumber
    Z_{QP} &= q_Q \left[ \sin{\beta_Q} K^{\eta}_{QP} \right. +(\cos{\beta_Q}/q^2_Q-\sin{\beta_Q} K^{\eta}_{QQ})  \\
    \label{eq-Zmat_and_Kreduced_f0g0}
    &\times (K^{\eta}_{QQ}+\tan{\beta_Q}/q^2_Q)^{-1} \left. K^{\eta}_{QP} \right] q_P,
\end{align}
where we have labeled $K^{\eta}$ the reaction matrix implementing the Coulomb functions pair $\{f^0,g^0\}$. The resulting ${\mathcal{K}_{PP}}$ and $Z_{QP}$ are still plugged into the final form of $\mathcal{F}(r)$ (right hand side of Eq. (\ref{eq-channel_elimination_F})) with $\{f,g\}$.

\subsection{A second set of independent solutions}

Jungen and Ross recognize at this point that their theory requires two additional elements:  ${\it (i)}$ more linearly independent solutions, in particular at least one additional solution is needed for each included symmetry that has a dissociation channel energetically open;  ${\it (ii)}$  one or more solutions with a nonzero value of the wavefunction on the boundary $R_0$, so the independent solutions can describe whatever negative logarithmic derivative in $R$ at the boundary that the true solutions satisfy at the dissociative boundary.  

Let's first address item   ${\it (ii)}$  by simply repeating the calculation listed above, but for a different logarithmic derivative imposed at $R_0$, such as $b^{(x)}=-\frac{\partial \Psi}{\partial R} \Psi^{-1}|_{R=R_0} =0$, as one natural choice.  This will give identical math to what was identified above, but we will identify the key quantities for this second set of $N_P$ independent solutions with a superscript $(x)$, namely $\mathcal{K}^{(x)}_{PP},Z_{QP}^{(x)}$. We identify the first set of solutions with the superscript $(0)$.

\begin{figure}
    \centering
    \includegraphics[width=1.0\linewidth]{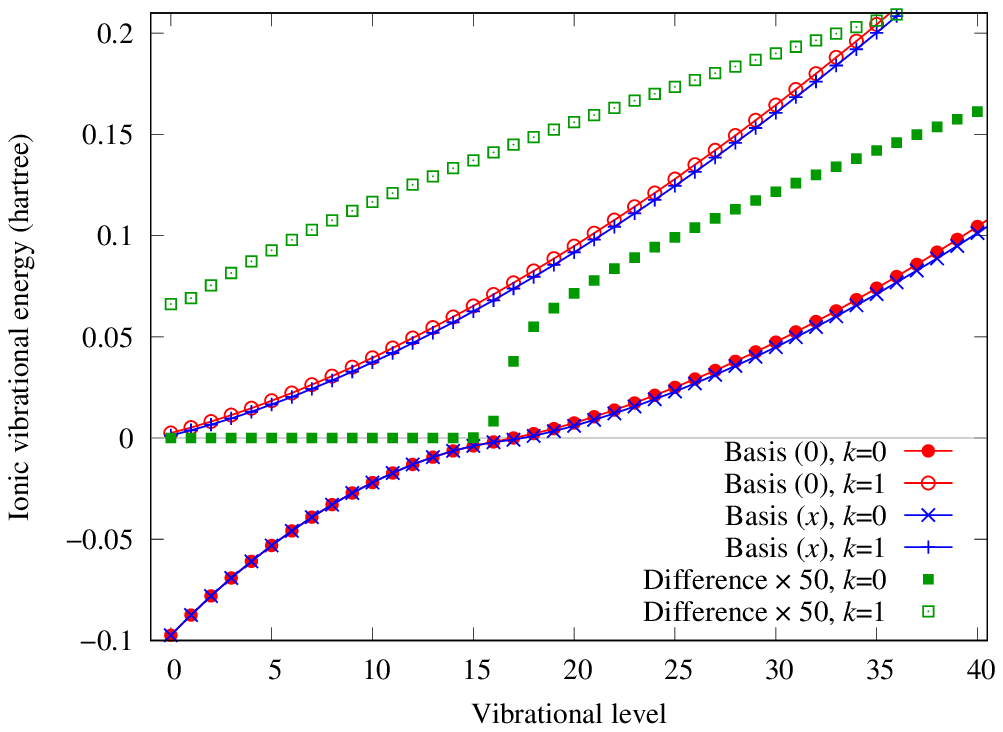}
    \caption{Vibrational energy levels in a.u. of H$_2^+$ versus level number $v$ for two different boundary conditions and both choices of the ionic potential curve $V^+_k$. The blue crosses represent the boundary condition ${\chi_{i}^{(x)}}{'}(R_0)=0$ and the red circles represent ${\chi_{i}^{(0)}}(R_0)=0$.  For this figure, $R_0=9$ a.u. The green squares show the differences (multiplied by 50) between them.  The development assumes that the vibrational eigenfunctions for the first $\sim$14 $k=0$ states are phase-standardized to be identical for the $(0)$ and $(x)$ solutions obeying different boundary conditions, i.e. $\chi_i^{(0)}(R) \approx \chi_i^{(x)}(R)$ for $v_i \lesssim 14$ and $R<R_0$.
    }
    \label{fig-vib_energies}
\end{figure}

To proceed, we first analyze the nature of the solutions obtained already in the vicinity of the boundary $R_0$, as a function of $r$.  Jungen and Ross point out that the nature of the dissociative solution(s) near the boundary can be identified in terms of the $Z$ coefficients discussed above.  Notice initially that only the $i \in Q$ have either a nonzero value or else a nonzero derivative on that boundary, so we can ignore the $i \in P$ portion of our vibrational wavefunction expansion, as we try to extract dissociative information from these independent solutions.
After this procedure is carried out, there will be two separate sets of $N_P$ independent solutions obeying different boundary conditions at $R_0$, and their electronically open/weakly-closed vibrational channel threshold energies are almost identical since the $i \in P$ subset of vibrational wavefunctions are entirely negligible in the vicinity of $R=R_0$. Fig. \ref{fig-vib_energies} compares the vibrational channel eigenenergies for the two different boundary conditions, namely vanishing wavefunction or vanishing $R$-derivative at $R_0=9$ a.u. for this example.
Note that for the $V^+_{k=0}$ ionic potential the blue set of energies corresponding to vanishing derivative at $R_0$ are nearly identical to the red set corresponding to vanishing vibrational wavefunction at $R_0$ for the lowest $\sim$14 $v$ states. For the other levels near energy 0 and higher, the red energy levels are higher since the vanishing wavefunction boundary condition is a tighter constraint than vanishing derivative.

The next steps are clearer if we write out more explicitly the wavefunctions from the two sets of solutions with the solution index $i'$ running from 1 to $N_P$ in both cases:
\begin{multline}
    \label{eq-wavefunction0}
    \Psi^{(K,0)}_{i'}=\sum_{i \in P} \ket{j_{i}} \chi_i^{(0)}(R) {\bigg (} f_i(r) \delta_{ii'} - g_i(r) {\cal K}_{ii'}^{(0)} {\bigg )} \\
    +\sum_{i \in Q} \chi_i^{(0)}(R)  W_i(r) Z_{ii'}^{(0)},
\end{multline}
with vanishing $\chi_i^{(0)}(R_0)=0$, and
\begin{multline}
    \label{eq-wavefunctionX}
    \Psi^{(K,x)}_{i'}=\sum_{i \in P} \ket{j_{i}} \chi_i^{(x)}(R) {\bigg (} f_i(r) \delta_{ii'} - g_i(r) {\cal K}_{ii'}^{(x)} {\bigg )} \\
    +\sum_{i \in Q} \chi_i^{(x)}(R)  W_i(r) Z_{ii'}^{(x)},
   \end{multline} 
with vanishing derivative, $\chi_i^{(x)}{'}(R_0)=0$.  Since the $\chi_i^{(0)}(R)$ and the  $\chi_i^{(x)}(R)$ are nearly identical for the lower vibrational levels $i \in P$, it will be desirable to treat them as identical for the purposes of the following calculations.  This will be a very good approximation.  Naturally, they are standardized to have the same phases, and in the following we can simply refer to both sets equivalently and omit the superscript for those channel functions having $i \in P$.  But one must keep in mind that the ionic vibrational functions are quite different for those high channel eigenfunctions with $i \in Q$.

At this point our approach deviates from the approach of Jungen and Ross \citet{Jungen_Ross_1997} (JR), but first consider how they proceed.  The remainder of their calculation is based on their equations (7)-(10).  The basic idea in those JR equations, which are specialized to the situation where each body-frame symmetry has only one open dissociation channel that is relevant, is to choose just \textbf{one }of the independent solutions out of the $N_P$ written for $\Psi^{(K,x)}_{i'}$ in Eq.~(\ref{eq-wavefunctionX}) above. Suppose we denote the one extra solution chosen as $i' = {i'}_x$, and it adds one more linearly independent solution (i.e. column vector in the sense of a solution matrix ${\underline F(r)}$)  to the set of solutions $\Psi^{(K,0)}_{i'}$.
Then JR add one more \textbf{row} to their solution matrix which represents the wavefunction on the dissociative surface $R_0$ in each of the $N_P+1$ linearly independent solutions in one symmetry (described here in \ref{subsection-extract_boundary}).  With these two steps JR effectively represent the full set of channels in this symmetry, namely $N_P$ ionization and their single included dissociation channel by a square reaction matrix $K_{q,q'}$, where $q$ is a composite index, $q'\equiv \{i', i'_x\}$, and part of the determination of that matrix  $K_{q,q'}$ is a symmetric orthogonalization of the full set of independent solutions to the Schr\"odinger equation.  The last step of the calculation is implementing the standard formulas of MQDT on a very fine energy mesh, to impose the exponentially decaying boundary conditions in all remaining ionization channels, using the standard channel elimination formula for Coulomb functions in the closed ionization channels ${i \in P}$.  That last step is basically the algebra written above in Eq. (\ref{eq-channel_elimination_S}), or its equivalent K-matrix variant restricted to the open-channel space.

\subsection{Combining the solution sets and extracting boundary information}
\subsubsection{Representation of the overcomplete solution set}
\label{subsection-overcomplete}
In the JR method there is some arbitrariness in selecting that one (for each symmetry) member $i'_x$ out of the full set of $N_P$ solutions $\Psi^{(K,x)}_{i'}$. While the additional needed independent solutions can be chosen in various possible ways, we describe here the method that we have found works the best in practice.  For starters, it is instructive to combine Eqs. (\ref{eq-wavefunction0}) and (\ref{eq-wavefunctionX}) while using a more generalized notation for the matrices on the right hand side. In the joined set of wavefunctions we use $B=0$ or $x$ to identify each subset, treating $B$ as an extension of the index $\{i',B\} \rightarrow i'_B$. Note again, that for $i \in P$ we can write $\chi_i^{(0)}(R) = \chi_i^{(x)}(R)$ (to an excellent approximation) but the strongly closed channel ionic states with very high $v$ quantum numbers are distinct for the two boundary conditions (now marked $i \in Q^{(B)}$). The joined set of wavefunctions is thus
\begin{multline}
  \label{eq-wavefunction_joined_1}
    \Psi^{(K,B)}_{i'}=\Psi^{(K)}_{i'_B}= \sum_{i \in P} \ket{j_{i}} \chi_i(R) {\bigg (} f_i(r) I_{i{i'_B}} -g_i(r) J_{i{i'_B}} {\bigg )} \\
    + \sum_{i \in Q^{(B)}}  \ket{j_{i}} \chi_i^{(B)}(R)  W_i(r) 
    \mathcal{Z}_{i{i'_B}},
\end{multline}
where $B=0$ or $x$ and $I_{i{i'_B}}$, $J_{i{i'_B}}$ and $\mathcal{Z}_{i{i'_B}}$ are the generalized matrices on the right hand side. They are rectangular with $2\times N_P$ columns. $I_{i{i'_B}}$ and $J_{i{i'_B}}$ have $N_P$ rows and $\mathcal{Z}_{i{i'_B}}$ has $2\times N_Q$ ($Q^{(B)}$ are distinct!) rows. In the K-matrix form they are
\begin{align}
  \label{eq-Imatrix_rectangle}
    \underline{I} &=\left(
    \begin{array}{c}
        \delta_{PP}  \; , \; \delta_{PP}
    \end{array}
    \right), \\
  \label{eq-Jmatrix_rectangle}
    \underline{J} &=\left(
    \begin{array}{c}
        \mathcal{K}^{(0)}_{PP}  \; , \; \mathcal{K}^{(x)}_{PP}
    \end{array}
    \right),\\
  \label{eq-Zmatrix_rectangle}
    \underline{\mathcal{Z}} &=\left(
    \begin{array}{cc}
        Z_{QP}^{(0)} & 0 \\
        0 & Z_{QP}^{(x)}
    \end{array}
    \right),
\end{align}
and they can easily be recast for other wavefunction forms. For example, in the sine--cosine form $\{C,S\}$ analogous to Eq. (\ref{eq-EDFT_psi_out}), $I$ consists of two channel-eliminated ${\cal S}^{(B)}_{PP}$ matrices, $J$ consists of two channel-eliminated ${\cal C}^{(B)}_{PP}$ matrices and $\mathcal{Z}$ merely has an altered prescription for the $Z_{QP}^{(0)}$ and $Z_{QP}^{(x)}$ blocks. The following omits writing the $(K)$ superscript in $\Psi_{i'_B}^{(K)}$, since all of the subsequent equations are equally valid for any solution set (e.g. the $K$-matrix form, or the $\{C,S\}$ form, for instance) that are written in terms of the real Coulomb functions. This is because the choice of the asymptotic normalization is now hidden inside the definition of the $\underline{I}$, $\underline{J}$, and $\underline{\mathcal{Z}}$ matrices in Eqs.~(\ref{eq-Imatrix_rectangle})--(\ref{eq-Zmatrix_rectangle}).

Before we find the correct linear combinations of this over-complete set of solutions, it is necessary to add extra rows to the matrices $\underline{I}$ and $\underline{J}$ corresponding to dissociation channels, e.g. the $EF$ and $GK$ states. We do this by analyzing the behavior of the solutions at the boundary $R=R_0$.

\subsubsection{Extraction of boundary information}
\label{subsection-extract_boundary}
Assuming the Born-Oppenheimer approximation holds at $R=R_0$, for molecular dissociation into $N_D$ distinct channels there will exist $N_D$ Born-Oppenheimer electronic wavefunctions $\ket{\phi_d^{\text{BO}}(r;R)}$ on the boundary $R=R_0$ satisfying Eq. (\ref{eq-BO_functions}).
The electronic part of a wavefunction dissociating into channel $d$ is this $\ket{\phi_{d}^{\text{BO}}(r;R)}$. Let us also denote the full set of dissociation channels $D$.

Recall that each ${i'_B}$ in Eq.~(\ref{eq-wavefunction_joined_1}) represents a separate solution of the Schr\"odinger equation with no boundary conditions imposed yet at $r \rightarrow \infty$ in the set of $N_P$ ionization channels. These solutions should, however, include a part of the full wavefunction near the dissociation surface $R=R_0$ which is the part bearing information about the molecular dissociation. It is expected that this part is represented by the second sum of Eq.~(\ref{eq-wavefunction_joined_1}) as the open and weakly closed vibrational channel functions $i \in P$ are all negligible at $R=R_0$. Therefore, any solution $\Psi_{i'_B}$ near $R \approx R_0$ can be expressed as
\begin{multline}
    \label{eq-match_joined}
    \left. \Psi_{i'_B}(r,R) \right |_{R=R_0} =
    \sum_{i \in Q^{(B)}} \left. \ket{j_i} \chi_i^{(B)}(R) \right |_{R=R_0}  W_i(r) 
    \mathcal{Z}_{i{i'_B}} \\
    \approx \sum_d \left. \ket{\phi_{d}^{\text{BO}}(r;R)} P_{di'_B}(E,R)\right |_{R=R_0} ,
\end{multline}
and neglecting the $R$-derivatives of the electronic wavefunctions in Eq.~(\ref{eq-match_joined}), in the spirit of the Born-Oppenheimer approximation, gives another equation:
\begin{multline}
    \label{eq-match_joinedX}
    \left. \frac{\partial}{\partial R} \Psi_{i'_B}(r,R) \right |_{R=R_0} \!\!\!\!=
    \sum_{i \in Q^{(B)}} \left. \ket{j_i} \frac{d}{dR} \chi_i^{(B)}(R) \right |_{R_0} \!\!\!\! W_i(r) \mathcal{Z}_{i{i'_B}} \\
    \approx \sum_d \ket{\phi_{d}^{\text{BO}}(r;R)} 
    \left. \frac{d}{dR} P_{di'_B}(E,R) \right |_{R=R_0} .
\end{multline}
Observe that for the basis $(0)$ the values of $\chi_i^{(B)}(R)$ are zero at $R_0$ and the $R$-derivative Eq. (\ref{eq-match_joinedX}) is the quantity of interest to be matched. Conversely, for basis $(x)$ the $R$-derivatives are always zero at $R_0$. In numerical tests, it is important to verify that the set of basis functions $\ket{\phi_{d}^{\text{BO}}(r;R)}$ are complete enough for the approximation in Eq. (\ref{eq-match_joined}) to be accurate.
In a more general case, without a convenient solvable 2D model like we implement in the present article, we might not have the Born-Oppenheimer electronic functions $\ket{\phi_d^{\text{BO}}(r;R)}$. However, we can still determine the bound state energies $U_d(R)$ and the asymptotic behavior of the corresponding electronic wavefunctions directly from the QDT data of Eq. (\ref{eq-QDF_matrix}) (regardless of how it is obtained). This alteration can be found in Appendix \ref{appendix-generalphi}

If we demand that $P_{di'_B}(E,R_0)$ should locally have an energy-normalized amplitude, this gives
\begin{equation}
\label{eq-dissociative_rows}
     P_{di'_B}(E,R_0) = {\bar F}_{E_d}(R_0) I_{d i'_B}-{\bar G}_{E_d}(R_0) J_{di'_B},
\end{equation}
where the functions $\{{\bar F}_{E_d}(R),{\bar G}_{E_d}(R)\}$ are a pair of independent radial solutions in the dissociative Born-Oppenheimer potential curve $U_d(R)$ and $E_d = E - U_d(R_0)$. These can be flexibly chosen to obey Milne-type \cite{Greene_Rau_Fano_1982,SeatonPeach1962,YooGreene1986} or WKB boundary conditions at $R_0$. Evidently, those channel contributions to a solution $\Psi_{i'_B}$ that are strongly-closed electronically will determine the contribution to dissociative channels on the $R=R_0$ surface.  These will thus determine the $I_{d i'_B}$ and $J_{d i'_B}$ elements of Eq. (\ref{eq-dissociative_rows}) that fill the needed extra rows in the full $\underline{I}$, $\underline{J}$ matrices, corresponding to dissociation channels. We extract them from $P_{di'_B}(E,R_0)$ using
\begin{align}
    \nonumber
    I_{d i'_B} = \frac{\mathcal{W}[P_{di'_B},{\bar G}_{E_d}]}{\mathcal{W}[{\bar F}_{E_d},{\bar G}_{E_d}]}, \\
    \label{eq-dissociation_IJ}
    J_{d i'_B} = \frac{\mathcal{W}[P_{di'_B},{\bar F}_{E_d}]}{\mathcal{W}[{\bar F}_{E_d},{\bar G}_{E_d}]},
\end{align}
where $\mathcal{W}[\cdot,\cdot]$ is a Wronskian in the $R$-coordinate evaluated at $R_0$. We can now express the $i'_B$-th solution as
\begin{multline}
  \label{eq-wavefunction_joined_full}
    \Psi_{i'_B}= \sum_{i \in P} \ket{j_i} \chi_i(R) {\bigg (} f_i(r) I_{i{i'_B}} -g_i(r) J_{i{i'_B}} {\bigg )} \\
    + \sum_d \ket{\phi_{d}^{\text{BO}}(r;R_0)} {\bigg (} {\bar F}_{E_d}(R) I_{d i'_B}-{\bar G}_{E_d}(R) J_{di'_B} {\bigg )}.
\end{multline}
For combining the ionization $i \in P$ and dissociation $d \in D$ channels, we can define a combined row index $\kappa \in \{P,D\}$.

\subsubsection{Construction of the full K matrix that includes both ionization and dissociation}
\label{subsection-fullK}
The expanded $I$, $J$ matrices have $N_P+N_D$ rows and $N_P\times 2$ columns (we assume $N_P > N_D$) due to employment of the two full sets of solutions (\ref{eq-wavefunction0}) and (\ref{eq-wavefunctionX}). We can now analyze the various approaches to reduction of their  columns to obtain dissociative solutions. The original JR method uses just one arbitrarily selected solution per extra basis. For example, we could try to emulate this somewhat by arbitrarily picking out $N_D$ solutions from the basis $(x)$ (perhaps picking those that have sufficiently different channel energies).
A much more accurate approach turns out to be choosing specific linear combinations of all of the solutions $\Psi_{i'_B}$ to construct a new set of solutions 
\begin{equation}
\label{eq-lin_comb_psi}
    \Psi_{\xi}=\sum_{i'_B} \Psi_{i'_B} C_{i'_B \xi} ,
\end{equation}
with some appropriately chosen linear combination coefficients $C_{i'_B \xi}$ with $N_P + N_D$ distinct indices $\xi$ (These $\xi$ do {\bf not} correspond to specific ionization or dissociation channels). This results in updated square $\underline{\mathcal{I}}$, $\underline{\mathcal{J}}$ matrices of the form
\begin{align}
    \nonumber
    \mathcal{I}_{\kappa \xi} &= \sum_{i'_B} I_{\kappa i'_B} C_{i'_B \xi},  \\
    \label{eq-lin_comb_IJ}
    \mathcal{J}_{\kappa \xi} &= \sum_{i'_B} J_{\kappa i'_B} C_{i'_B \xi},
\end{align}
for all channels $\kappa \in \{ P, D \}$. Then, it is a simple matter to construct the intermediate K matrix $\underline{\bar{K}}^{\text{int}}=\underline{\mathcal{J}} \underline{\mathcal{I}}^{-1}$ and write the combined solutions
\begin{multline}
  \label{eq-wavefunction_combined_1}
    \Psi_{\kappa'}= \sum_{i \in P} \ket{j_i} \chi_i(R) {\bigg (} f_i(r) \delta_{i\kappa'} -g_i(r) \bar{K}^{\text{int}}_{i\kappa'} {\bigg )} \\
    + \sum_d \ket{\phi_{d}^{\text{BO}}(r)} {\bigg (} {\bar F}_{E_d}(R) \delta_{d\kappa'} - {\bar G}_{E_d}(R) \bar{K}^{\text{int}}_{d\kappa'} {\bigg )},
\end{multline}
where $\kappa'\in \{ P, D \}$ now do correspond to the ionization and dissociation channels.

In the course of the present study, a number of alternative approaches have been implemented to combine the columns of the matrices $\underline{I}$, $\underline{J}$. 
Only our most successful approach is presented here, motivated purely mathematically:
In short, we simply combine all of the $2\times N_P$ solutions to create a set of $N_P+N_D$ solutions having the maximal linear independence. This can be done, for example, using the singular value decomposition of matrix $\underline{J}$
\begin{equation}
    \label{eq-SVD_of_J}
    \underline{J} = \underline{\mathcal{U}} \; \underline{\Sigma} \; \underline{\mathcal{V}}^{\dagger} ,
\end{equation}
where $\underline{\mathcal{U}}$ is an $(N_P+N_D)\times(N_P+N_D)$ square unitary matrix, $\underline{\mathcal{V}}$ is a $(2N_P)\times(2N_P)$ square unitary matrix and $\underline{\Sigma}$ is an $(N_P+N_D)\times(2N_P)$ rectangular diagonal matrix (for a real $\underline{J}$, these three matrices are real).
Denoting by $\Sigma_{\xi}$ the non-zero singular values from matrix $\underline{\Sigma}$, and noting that each has a corresponding row in the matrix $\underline{\mathcal{V}}^{\dagger}$ (or column in $\underline{\mathcal{V}}$), 
we can then create
\begin{equation}
\label{eq-lin_comb_SVD}
    C_{i'_B \xi} = \mathcal{V}_{i'_B \xi} / \Sigma_{\xi},
\end{equation}
for all $\xi$. The number of non-zero singular values needs of course to equal $N_P+N_D$, in order to yield the correct number of independent solutions. Note that now  $\underline{\mathcal{J}} = \underline{\mathcal{U}}$, which is a  linearly-independent set of columns. We can rewrite the computation of $\underline{\bar{K}}^{\text{int}}$ as
\begin{equation}
\label{eq-K_is_JIinv}
    \underline{\bar K}^{\text{int}} = 
    \left( \underline{\mathcal{I}} \; \underline{\mathcal{J}}^{-1} \right)^{-1} =
    \left( \underline{I} \; \underline{\mathcal{V}} \; \underline{\Sigma}^{-1} \underline{\mathcal{U}}^{\dagger} \right)^{-1},
\end{equation}
where the rectangular diagonal matrix $\underline{\Sigma}^{-1}$ is the matrix $\underline{\Sigma}$ with its elements inverted and transposed. The keen reader will note that the term $\underline{\mathcal{V}}\; \underline{\Sigma}^{-1} \underline{\mathcal{U}}^{\dagger}$ is known as the Moore-Penrose inverse (or pseudoinverse) of the rectangular matrix $\underline{J}$. This approach can be analogously derived from the singular value decomposition of the matrix $\underline{I}$ and its ensuing pseudoinverse.

Once a reaction matrix $\underline{\bar{K}}^{\text{int}}$ is determined which is based on the use of the $\{{\bar F}_E(R),{\bar G}_E(R)\}$ pair in one or more open dissociative channels, it is straightforward to transform it to refer to standard physical solutions $\{{ F}_E(R),{ G}_E(R)\}$ with energy normalized amplitudes at $R\rightarrow \infty$ and asymptotic phase differing by exactly $\pi/2$. Using the Wronskian identity that holds among any three solutions to the radial Schr\"odinger equation, e.g. ${\bar F} {\cal W}[F,G]+F {\cal W}[G,{\bar F}]+G{\cal W}[{\bar F},F]=0$ allows us to express $\{{\bar F}_E(R),{\bar G}_E(R)\}$ in terms of $\{F_E(R),G_E(R)\}$ to obtain the more physical (but still containing closed channels) matrix $\underline{K}^{\text{int}}$.
Similarly, if a different normalization had been used for the Coulomb functions in the ionization channels, then at this point the transformation is made to express the reaction matrix or equivalent scattering information in terms of the standard energy-normalized Coulomb functions.

From here, everything proceeds as in ordinary MQDT. The real, symmetric reaction matrix $\underline{K}^{\text{int}}$ still contains closed ionization channels (and possibly closed dissociation channels). The channel elimination formula of Eq. (\ref{eq-channel_elimination_K_basic2}) can be implemented on a very fine energy mesh to enforce exponential decay in all remaining closed channels and determine the physical reaction matrix, $\underline{K}^{\rm phys}$.
If closed dissociation channels are also included with Milne-type asymptotic behavior, the closed channel elimination formula needs to be slightly adjusted, precisely as described in our previous H$_2^+$ gerade publication, appendix A \cite{EPJD_H2_Gerade_2022}.  The preceding sentence assumes that one would normally make a Born-Oppenheimer approximation when computing each solution (column vector) at $R>R_0$, which matches smoothly (continuous in function and derivative) to the solution and $R$-derivative determined in the above procedure at $R_0$.  But if nonadiabatic corrections or avoided crossings are present in the long range Born-Oppenheimer potential curves, those phenomena are readily described using conventional coupled equations or else approximately using Landau-Zener-St\"uckelberg theory.  Afterwards, the physical scattering matrix will of course be
\begin{equation}
    \underline{S}^{\rm phys} = \frac{\underline{1}+i \underline{K}^{\rm phys}}{\underline{1}-i\underline{K}^{\rm phys}}
\end{equation}
And from this point, cross sections for vibrational excitation and dissociative recombination utilize the usual formulas
\cite{Curik_HG_2DRmat_2018}.

\subsection{Computational details}
Most of our computations in the following section are done with box sizes $r_0 = 6$~bohr and $R_0=9$~bohr, though for the higher energy regions we increase $R_0$ to 11 bohr. We study scattering energies up to the second vibrational excitation threshold of the $1s\sigma$ ionic potential.
To reach a convergence with respect to the vibrational basis we use 80 basis functions for $j=0,2$ and 85 for $j=1$. In the case of the 11 bohr nuclear box and higher scattering energies, these numbers need to be raised to about 130. The number of open and weakly-closed channels $N_P$ is 12, split evenly between $j=0$ and $j=2$ channels. All of the $j=1$ channels are treated as strongly closed. The number $N_P$ can be varied without any substantial impact on the final results as long as it's not too large such that the $\chi_i^{(0)}(R)$ and $\chi_i^{(x)}(R)$ do not differ appreciably (and also $N_P \geq N_D$ must hold).
In the range of energies studied there are three open dissociation channels. These are channels connected to the $EF$ and $GK$ potential curves with asymptotic energy of $1/8$~a.u. below the ionization threshold and to the $X$ curve with asymptotic energy of $1/2$~a.u. below the ionization threshold. However, our model Hamiltonian reproduces the $X$ potential curve very inaccurately and its corresponding DR cross section tends to be several orders of magnitude below all others. Therefore, while the ground-state curve $X$ is present in the calculations, its corresponding cross sections are not analysed here.

To complete the description of closed-channel resonances
it is also necessary to include the $H\bar{H}$ DR channel, which is asymptotically closed but open locally at short internuclear distances. Including the higher channels linked to the $O$ and $P$ potential curves makes no appreciable difference in the studied range of energies.
Solutions are matched to Milne-type asymptotic functions \cite{Greene_Rau_Fano_1982,SeatonPeach1962,YooGreene1986} 
for all dissociating channels except $X$ in Eq.~(\ref{eq-dissociation_IJ}).  The need to treat the dissociative coordinate motion in the outer region $R>R_0$ is two-fold.  First, it ensures that the near-threshold scattering matrices have the correct Wigner-threshold law behavior at every dissociation threshold.  Secondly, there can be high-lying vibrational bound states in potentials such as the $H\bar{H}$ potential which extend far beyond $R_0$ and our use of the Milne variant of generalized QDT ensures that those high-lying vibrational bound states are described accurately within the Born-Oppenheimer approximation.  In order to treat some phenomena such as resonant ion-pair formation, very long range non-Born-Oppenheimer couplings can become important. as in the well-known crossing near $R=35$ a.u. that is treated in detail in Ref.\cite{hornquist2024dissociative}.  In such cases one must solve the coupled dissociative equations in the region $R>R_0$ in order to describe all the relevant physics correctly including ion-pair formation.

An additional key aspect of the present theoretical description should be made clear at this point: The scattering information needed for a full implementation {\it requires} that the body-frame scattering data, i.e. the underlying quantity $K_{jj'}({\cal E},R)$ that the entire theory is based on, is needed at both positive (open) and {\it negative (closed)} channel energies.  Some scattering codes provide negative or closed-channel energy collision matrices, notably the fixed-nuclei UK Molecular $R$-matrix Codes \cite{MorganRm_CPC,TennysonUK} that can be connected with MQDT solutions in the outer region as was carried out in Refs.\cite{hvizdovs2023bound, ForerCF}. But many frequently-used programs such as the complex Kohn variational program\cite{schneider1988complex} or the Schwinger variational principle code\cite{lucchese1986applications} do not routinely (or ever) produce scattering matrices at negative electron collision energies that include weakly-closed channels in the MQDT sense.  In order for body-frame calculations to utilize our ideas presented here, it would be crucial to extend those codes to negative scattering energies, as has been done in at least one instance for atoms \cite{goforth1992multichannel}. In the present treatment, for instance, body-frame energies ${\cal E}$ required for converged calculations can extend over energies from as low as -0.5 a.u. up to 0.3 a.u. near $R=2$ a.u., which translates approximately to body-frame channel energies $\varepsilon_{j=0,2}(R)=$-0.4 a.u. to 0.4 a.u. and $\varepsilon_{j=1}(R)=$-0.8 a.u. to 0 a.u.

\section{Cross section comparisons }

Results from our modified version of the Jungen-Ross treatment to the model H$_2^+ + e^-$ gerade system, computed using energy-dependent frame transformation theory, are compared here with the vibrational excitation and dissociative recombination results obtained using the 2D $R$-matrix approach \cite{Curik_HG_2DRmat_2018}.

\begin{figure}[tbh]
    \centering
    \includegraphics[width=1.0\linewidth]{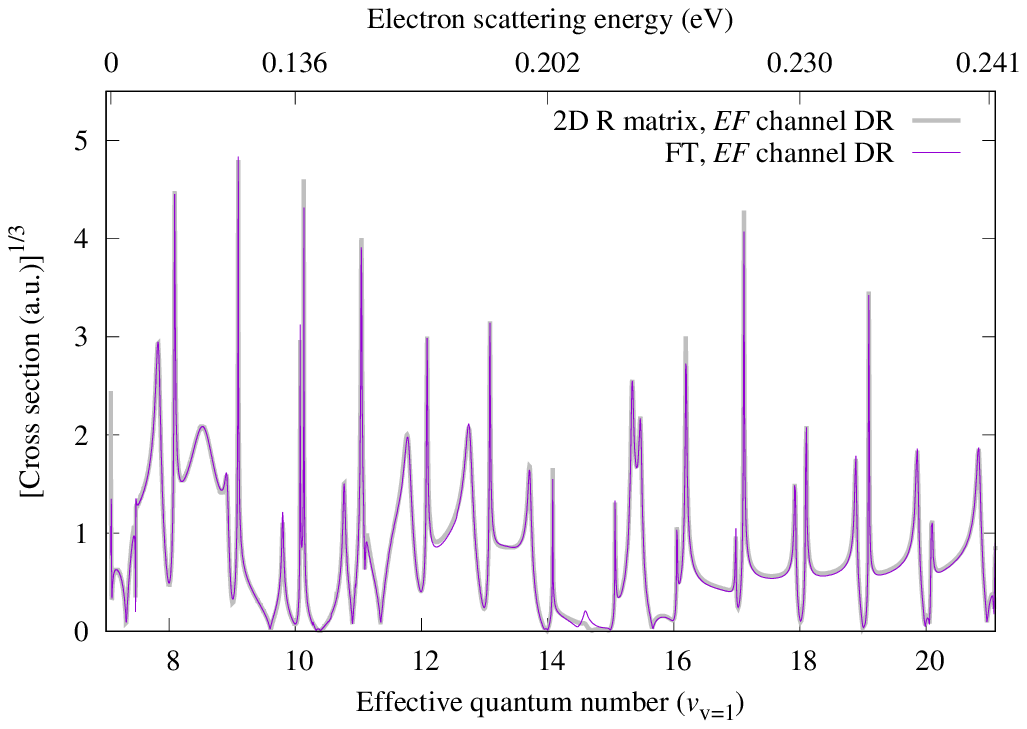}
    \caption{Comparison of partial dissociative recombination cross sections. The input channel is $i=\{v=0,j=0\}$ and the output channel is the $EF$ curve. The incoming electron energies span from zero to $(E_{1,0}-0.031\text{ eV})$.
    }
    \label{fig-DR_1a}
\end{figure}

\begin{figure}[tbh]
    \centering
    \includegraphics[width=1.0\linewidth]{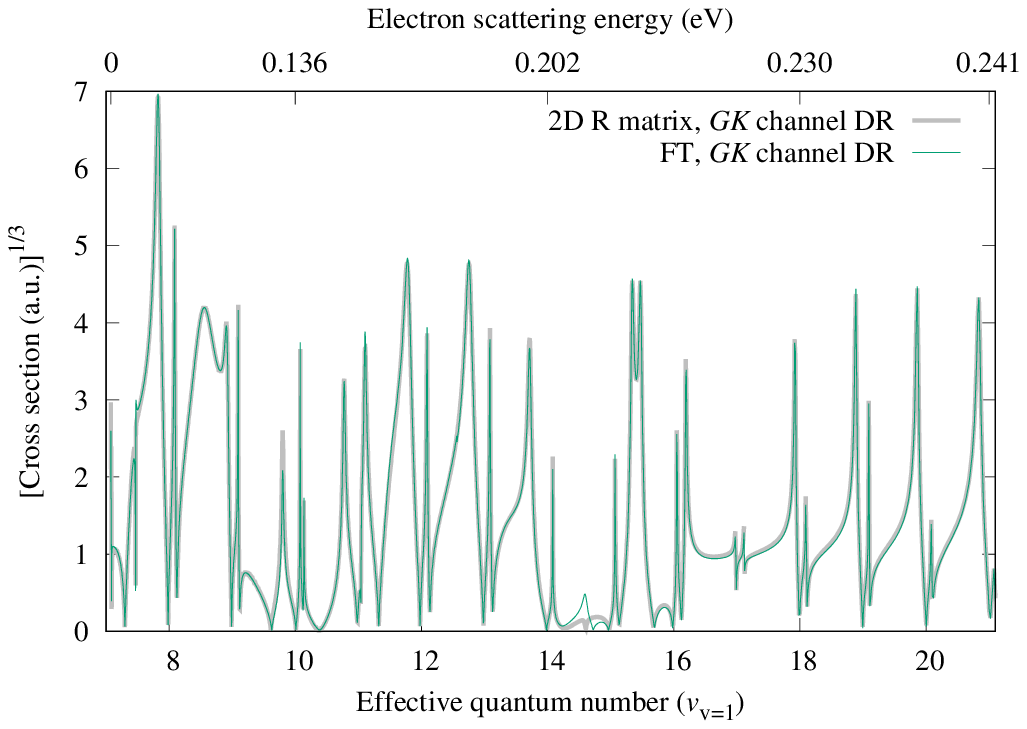}
    \caption{Comparison of partial dissociative recombination cross sections. The input channel is $i=\{v=0,j=2\}$ and the output channel is the $GK$ curve. The incoming electron energies span from zero to $(E_{1,0}-0.031\text{ eV})$.
    }
    \label{fig-DR_1b}
\end{figure}

\begin{figure}[tbh]
    \centering
    \includegraphics[width=1.0\linewidth]{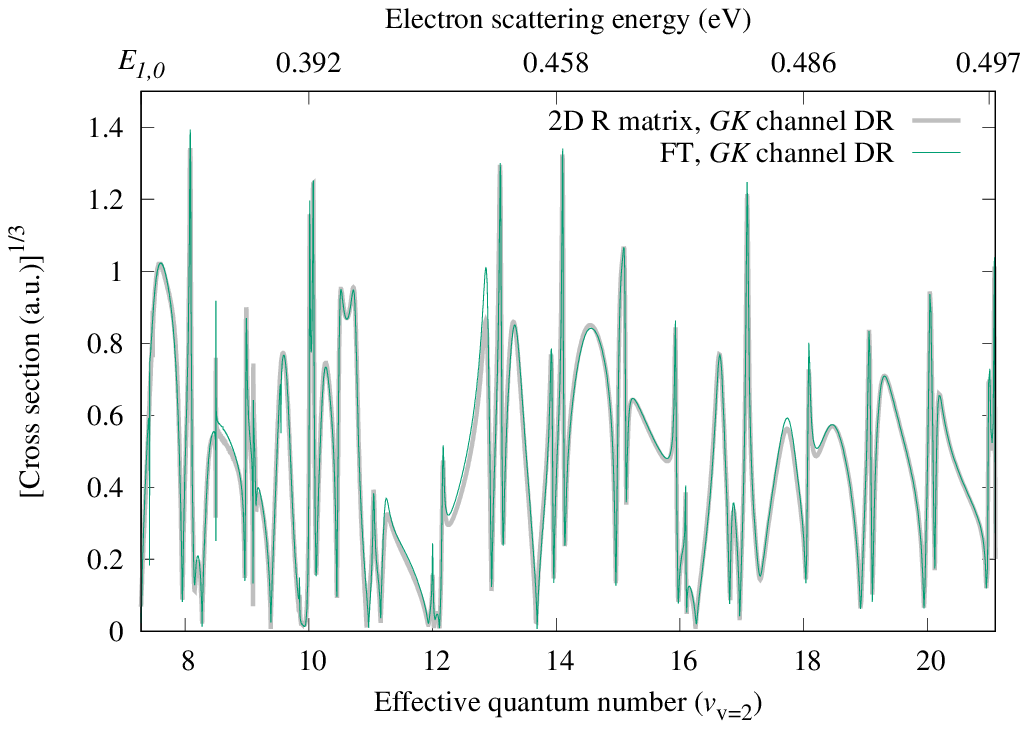}
        \caption{Comparison of partial dissociative recombination cross sections. The input channel is $i=\{v=0,j=0\}$ and the output channel is the $GK$ curve. The incoming electron energies span from the first excitation threshold $E_{1,0}$ to $(E_{2,0}-0.031\text{ eV})$.
        }
    \label{fig-DR_2a}
\end{figure}

\begin{figure}[tbh]
    \centering
    \includegraphics[width=1.0\linewidth]{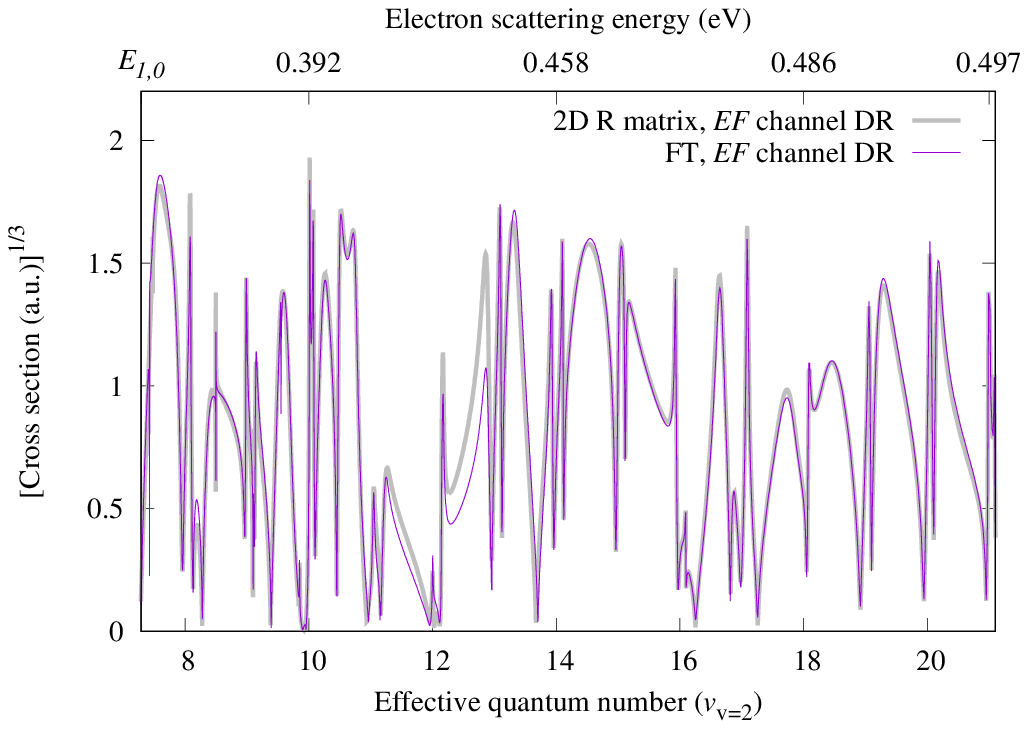}
        \caption{Comparison of partial dissociative recombination cross sections, plotted on a cube root scale versus effective quantum number, in order to improve visibility of the details. The input channel is $i=\{v=0,j=2\}$ and the output channel is the $EF$ curve. The incoming electron energies span from the first excitation threshold $E_{1,0}$ to $(E_{2,0}-0.031\text{ eV})$.
        }
    \label{fig-DR_2b}
\end{figure}

Figs.~\ref{fig-DR_1a}-\ref{fig-DR_2b} show the comparison for partial dissociative recombination cross sections into $EF$ or $GK$ output channels.
The input channels are the ground vibrational states $v=0$ for either $j=0$ or $j=2$ (these have the same nuclear vibrational wavefunction $\chi_i(R)$ but different electronic angular momentum). We show two of these input-output combinations in the energy window from zero scattering energy to just below the first vibrational excitation threshold and two more at energies between the first and second excitation threshold.

Some of the strong features apparent in Figs.~\ref{fig-DR_1a}-\ref{fig-DR_2b} are complex resonances of the type found in LiH$^+$ DR to play an especially strong role in creating enhanced DR rates.\cite{CurikLiH2007prl}  One difference here from that LiH system is that the DR in LiH was totally dominated at low energy by indirect dissociation since there is no direct DR pathway.  The gerade H$_2$ system of course has a strong direct pathway, which creates relatively strong recombination cross sections even away from those complex resonances.  Nevertheless, strong enhancements at complex Rydberg resonances do appear throughout the DR figures shown here, such as the region from $\nu_2=16-20$ in Fig.\ref{fig-DR_2b} as one of the clearest examples of the phenomenon.  Recall that these types of complex resonances are effective in creating strong DR probabilities because when multiple fine Rydberg resonances having relatively high $n$ and low $v$ are spread across a broad lower $n$ and higher $v$ resonance, this interchannel coupling aids in exciting the motion of the heavy nuclei.

\begin{figure}[ht]
    \centering
    \includegraphics[width=1.0\linewidth]{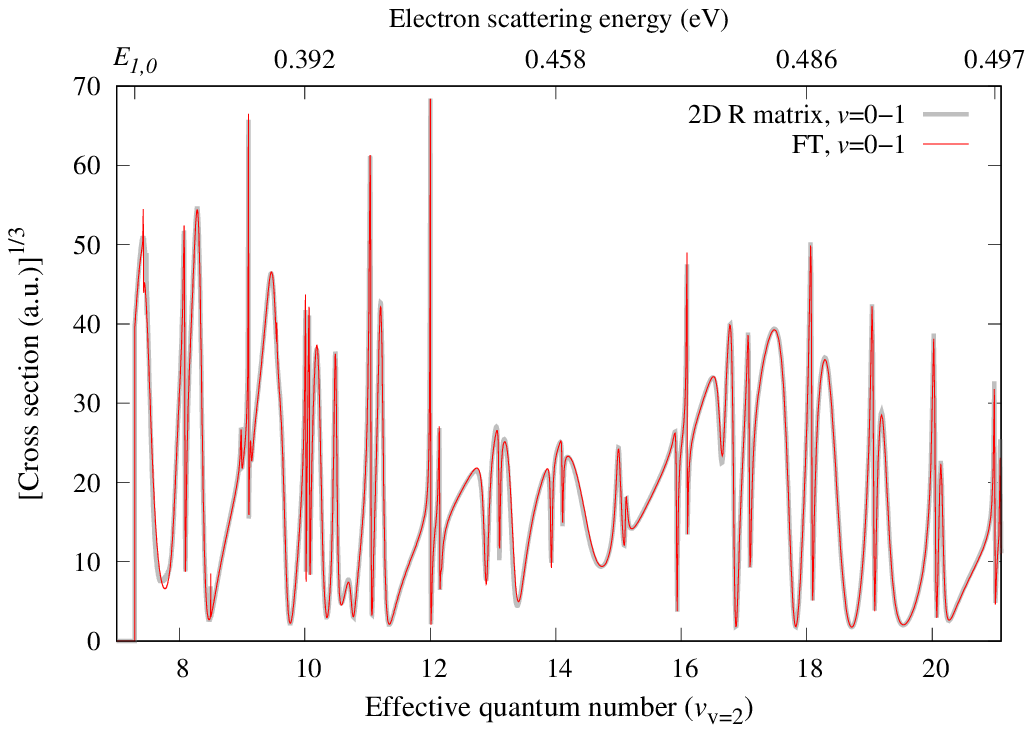}
    \caption{Comparison of $v=0\rightarrow 1$ vibrational excitation cross sections. The curves are a sum over all four combinations of input and output $j$ values. The incoming electron energies span from the first excitation threshold $E_{1,0}$ to $(E_{2,0}-0.031\text{ eV})$.
    }
    \label{fig-VE_2}
\end{figure}

\begin{figure}[ht]
    \centering
    \includegraphics[width=1.0\linewidth]{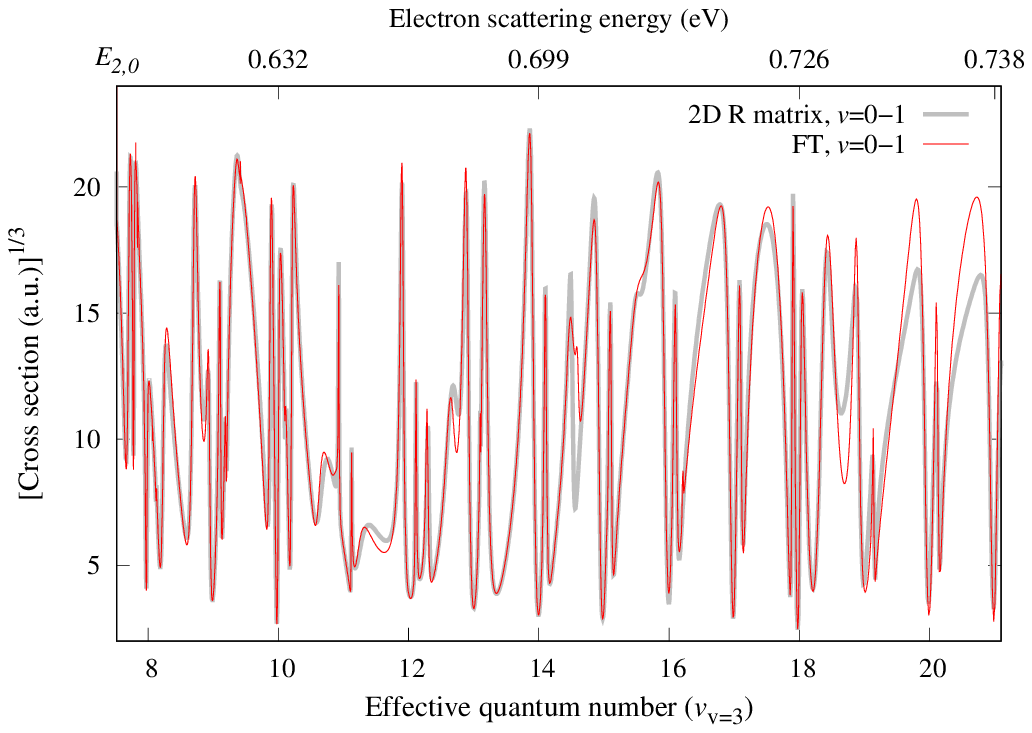}
    \caption{Comparison of $v=0\rightarrow 1$ vibrational excitation cross sections. The curves are a sum over all four combinations of input and output $j$. The incoming electron energies span from the first excitation threshold $E_{2,0}$ to $(E_{3,0}-0.031\text{ eV})$.
    }
    \label{fig-VE_3a}
\end{figure}

\begin{figure}[ht]
    \centering
    \includegraphics[width=1.0\linewidth]{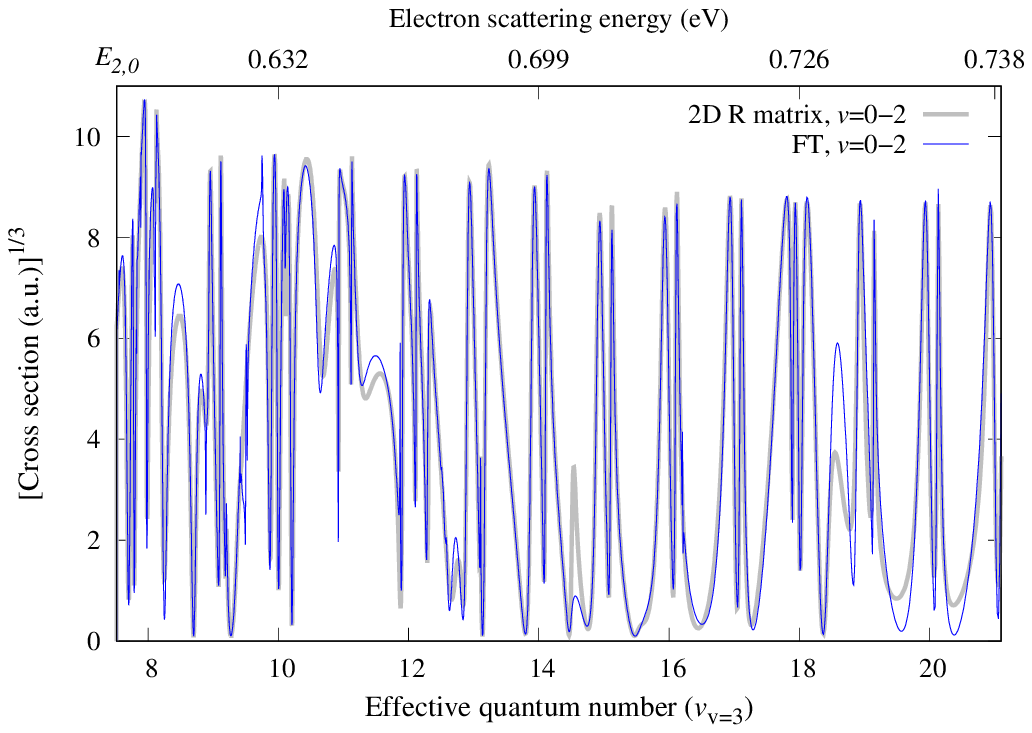}
    \caption{Comparison of $v=0\rightarrow 2$ vibrational excitation cross sections. The curves are a sum over all four combinations of input and output $j$. The incoming electron energies span from the first excitation threshold $E_{2,0}$ to $(E_{3,0}-0.031\text{ eV})$.
    }
    \label{fig-VE_3b}
\end{figure}

The Figs.~\ref{fig-VE_2}-\ref{fig-VE_3b} compare cross sections of vibrational excitation into the first and second excited vibrational states. These cross sections are summed over all combinations of input and output $j$ and are computed in the energy windows between the first and second excitation threshold and the second and third excitation threshold. The cross section summed over all the input and output channels can also be interpreted as the laboratory-frame cross section obtained by an average over all the molecular orientations \cite{Gianturco_inbook}.

The disagreements between the FT and exact results mostly appear as very narrow "glitches" where the FT approach seems to pick up an extra resonance, though sometimes we also see broader erroneous features. The glitch errors are usually correlated with the fitting error of the dissociation surface matching of Eq. (\ref{eq-match_joined}) or numerical errors produced from computing $\underline{\bar{K}}^{\text{int}} = \underline{\mathcal{J}} \underline{\mathcal{I}}^{-1}$ around poles of $\underline{J}$ or $\underline{I}$.
All of these errors are of a numerical nature and as such appear at rather random energies. As such, it is often possible to get rid of them by simply repeating the computation with slightly changed computational parameters (e.g. number of vibrational states, electronic box size $r_0$, inversion of $\underline{\mathcal{I}}$ or $\underline{\mathcal{J}}$ in the $\underline{\bar{K}}^{\text{int}}$ computation) and comparing the results. For this we simply need to know at which energies the errors occur. Fortunately, this can be diagnosed simply by looking at the eigendefect sum obtained by diagonalizing the final reaction matrix $\underline{K}^{\rm phys}$. This eigendefect sum typically exhibits wild unphysical behavior such as rapidly plunging values around the aforementioned errors.


\section{Conclusions}
The present study has introduced a generalization of the energy-dependent vibrational frame transformation. In combination with multichannel quantum defect theory, the theory can describe competing bound Rydberg states, ionization channels, and dissociation channels and produce accurate scattering matrices containing all possible ways the molecular system can fragment.  The greatest novelty is demonstrating that one can now use this theory 
to map the body-frame scattering information, which depends on the body-frame energy ${\cal E}$ and the internuclear distance $R$, into a physical scattering matrix $S(E)$ that varies strongly and resonantly with the total laboratory frame energy $E$.  Moreover, this theory has been demonstrated here to work accurately for our benchmarked realistic model of the $^1\Sigma_g$ symmetry of H$_2$; that model can be solved essentially exactly, even when multiple ionic potential energy curves exist that support a rich complexity of corresponding Rydberg and ionization and dissociation channels.  A key component is our adaptation and extension of the powerful formulation of Jungen and Ross,\cite{Jungen_Ross_1997} which they developed to describe competing ionization and dissociation channels, and which they applied successfully to H$_2$ using their own model of the electron-H$_2^+$ energy-independent body-frame reaction matrices.  This allows us to create a highly accurate mapping of ${\cal E}$-dependent fixed-nuclei body-frame scattering information into the physical scattering matrix that can now treat processes such as dissociative recombination with spectroscopic accuracy in the computed resonance positions, widths, and lineshapes.  For the moment, our focus has been exclusively on the vibrational frame transformation, but additional steps can be introduced in the usual manner, such as the rotational frame transformation, and as needed, the incorporation of fine or hyperfine structure.\cite{Jungen_Dill_JCP_1980, OsterwalderMerktJungen2004}

One future application of the theory which remains desirable is its application to the {\it ab initio} body-frame data that can be computed for electron scattering from H$_2^+$ and its isotopologues, e.g. using $R$-matrix methods to solve the fixed-nuclei Schr\"odinger equation with the true Hamiltonian.  Such body-frame calculations have been carried out previously, as in \cite{GreeneYoo1995, TelminiJungen2004}, although for a high accuracy treatment of dissociative recombination (DR) it will be advantageous to obtain still more precision than has been achieved to date (or else to introduce small, semi-empirical corrections that guarantee high accuracy in the Born-Oppenheimer potential curves).  The path of this theoretical trajectory appears to be poised to improve upon existing DR theory (see, e.g. \cite{waffeu2011assignment}) and provide a first principles theory that can even describe the low energy resonances of H$_2$ and its isotopologues with precision approaching the 1 cm$^{-1}$ level. Higher accuracy could also be achieved by implementing the methods of this paper in place of the energy independent frame transformation used in Refs. \cite{ForerCF,ForerCH}.

At higher energies, like in the range 2-10 eV, other channels are open, such as resonant ion-pair formation that occurs when an electron collides with HD$^+$, and this higher energy range (and larger $R$-range) would constitute a stringent test of the current theoretical treatment.  That process received experimental attention that could be interpreted  semi-quantitatively using Landau-Zener-St\"uckelberg theory,\cite{DunnLarson2000} and very recently a much more detailed theory treatment achieved good agreement with experiment.\cite{hornquist2024dissociative} Another desirable future extension would apply this formulation to the dissociative recombination of light polyatomic ions such as H$_3^+$.

\section{Acknowledgements}
We  thank Christian Jungen for discussions and advice.  The Purdue portion of this work was supported by the Department of Energy, Office of Science, Office of Basic Energy Sciences, Award No. SC0010545. R.\v{C}. acknowledges support of the Czech Science Foundation (Grant No.\ GACR 21-12598S).

\appendix

\section{\label{appendix-generalphi}Extraction of dissociating information from QDT data}
The QDT data of Eqs. (\ref{eq-QDF_matrix}) allows us to determine the values of $U_d(R_0)$ and the asymptotic behavior of the corresponding electronic functions $\ket{\phi_d(r;R_0)}$. For this derivation let us use the sine/cosine form of an unphysical solution determined by the matrices ${\cal S}_{jj'}({\cal E}, R_0)$ and ${\cal C}_{jj'}({\cal E}, R_0)$
\begin{multline}
\label{eq-phi_sincos}
    \ket{\phi_{j'}({\cal E},r;R_0)} \rightarrow \\ \sum_{j} \ket{j}
    \bigg( f_j(r) {\cal C}_{jj'}({\cal E},R_0) - g_j(r) {\cal S}_{jj'}({\cal E}, R_0) \bigg),
\end{multline}
in the $r \rightarrow \infty$ limit. In the bound-state region, all energies $\varepsilon_j={\cal E}-V^+_{k_j}(R_0)$ are negative. Physical solutions exist only at energies where all exponentially rising parts can be eliminated \cite{OrangeReview} i.e. whenever there exists a linear combination ${\cal A}_{j'}$ such that
\begin{equation}
\label{eq-Dmatrix}
    \sum_{j'} (\sin\beta_j {\cal C}_{jj'} + \cos\beta_j {\cal S}_{jj'}) {\cal A}_{j'} = 0,
\end{equation}
with $\beta_j$ defined in terms of the fixed-nucleus quantum numbers.
These energies give the $U_d(R_0)$ and we denote ${\cal A}_{j'd}$ and $\varepsilon_{jd}=U_d(R_0)-V^+_{k_j}(R_0)$ the corresponding quantities. The asymptotic behavior of the physical bound-state functions is then
\begin{multline}
\label{eq-general_phi_d}
    \ket{\phi_{d}(r;R_0)} \rightarrow \frac{1}{N_d}
    \sum_{j j'} \ket{j} \\
    \times \bigg( f_j(r) {\cal C}_{jj'}(U_d,R_0) 
    - g_j(r) {\cal S}_{jj'}(U_d, R_0) \bigg) {\cal A}_{j' d},
\end{multline}
with the Coulomb functions evaluated at energies $\varepsilon_{jd}$ and
where the normalization is derived by generalizing the steps of Ref. \cite{Greene_Fano_Strinati}
\begin{multline}
\label{eq-phi_normalization}
    |N_d|^2 = \sum_j \frac{1}{\pi} \left( \sum_{j'}(\cos\beta_j {\cal C}_{jj'} - \sin\beta_j {\cal S}_{jj'}){\cal A}_{j'd} \right) \\
    \times \left. \left( \sum_{j''} X'_{jj''} {\cal A}_{j''d} \right) \right|_{{\cal E}=U_d},
\end{multline}
with
\begin{multline}
\label{eq-phi_normX}
    X'_{jj''} = \frac{d \beta_j}{d {\cal E}} \left(\cos\beta_j {\cal C}_{jj''} - \sin\beta_j {\cal S}_{jj''} \right) + \\ 
    \left(\sin\beta_j \frac{d}{d {\cal E}} {\cal C}_{jj''} + \cos\beta_j \frac{d}{d {\cal E}} {\cal S}_{jj''} \right).
\end{multline}
%


\bibliography{bibliography,jungen2}
\end{document}